\documentclass{elsart}
\usepackage{epsfig,fleqn}
\newcommand{\ie}{{\it i.e.}\ }
\newcommand{\eg}{{\it e.g.}\ }
\newcommand{\cf}{{\it cf.}\ }
\newcommand{\be}{\begin{eqnarray}}
\newcommand{\ee}{\end{eqnarray}}

\newcommand{\vect}[1]{{\vec{#1}}}

\def\half{\textstyle{\frac{1}{2}}}
\def\quarter{{\textstyle{\frac{1}{4}}}}
\def\third{{\textstyle{\frac{1}{3}}}}

\def\mintp{\int\!\frac{d^4 p}{(2\pi)^4}}
\def\mintpi{\int\!\frac{d^4 p_i}{(2\pi)^4}}
\def\mintpf{\int\!\frac{d^4 p_f}{(2\pi)^4}}
\def\mintpp{\int\!\frac{d^4 p'}{(2\pi)^4}}

\def\L{\mathcal L}
\def\del{\partial}
\begin{document} 
\begin{frontmatter}
{\small
 ~  \hfill   UNITU-THEP-9/1997  \newline
 hep-ph/9705267       \hfill   TU-GK-97-003                     }

\title{Nucleon Form Factors in a Covariant Diquark--Quark Model$^\dag$}
\baselineskip=18 true pt
\author{
G.\ Hellstern, 
R.\ Alkofer,
M.\ Oettel
and H.\ Reinhardt} 
\address{Institute for Theoretical Physics, T\"ubingen University \\
Auf der Morgenstelle 14, D-72076 T\"ubingen, Germany}
%\maketitle

\vskip 1cm

\begin{abstract}
In a model where constituent quarks and diquarks interact through quark
exchange the Bethe--Salpeter equation in ladder approximation for the nucleon
is solved.  Quark and diquark confinement is effectively parametrized by 
choosing  appropriately modified propagators. The coupling to external 
currents is  implemented via nontrivial vertex functions  for quarks and 
diquarks to ensure gauge invariance at the constituent level.
Nucleon matrix elements are evaluated in a generalised impulse approximation,
and electromagnetic, pionic and axial form factors are calculated.  
\vskip 1cm
\end{abstract}
\end{frontmatter}

\vfill
\noindent
$^\dag$Supported by COSY under contract 41315266, BMBF under
contract 06TU888\\
and Graduiertenkolleg ''Hadronen und Kerne'' (DFG Mu705/3).
 
\eject

\section{Introduction}

Accelerators like ELSA, CEBAF and COSY will investigate hadron observables at
a scale intermediate to the low--energy region where hadron phenomena mainly
reflect the underlying principles of chiral symmetry and its dynamical
breakdown, and to the large--momentum regime where the ``strong'' interaction
has the appearance of being a perturbation on free--moving quarks and gluons.
A theoretical description of the corresponding intermediate energy physics
aims at an understanding of the interplay between hadronic degrees of freedom
and their intrinsic quark substructure.  
At present the description of the intermediate energy region 
of several \mbox{GeV}, however, 
requires to use an effective parametrization of
confinement within a fully relativistic formalism. Furthermore, in the
description of hadrons as bound states of quarks nonperturbative methods are
unavoidable. On the other hand, for large momentum transfers the perturbative
results of QCD should be met.  In this paper we set up a further step in the
development of an interpolating model to describe baryon structure in the
intermediate energy region.

The basic idea of this model is to parametrize complicated and/or unknown
structures within baryons by means of two--quark correlations in the overall
antisymmetrized color--antitriplet channels. This amounts to describing
baryons effectively as bound states of constituent quarks and diquarks (for a
recent digest of diquark models see \eg \cite{Diq3}).  Guided by the notion
that a fully relativistic Faddeev equation determines baryons as bound states
of three quarks and provides a complete and correct picture of baryons we
will study the Bethe--Salpeter equation for a diquark--quark system. The
minimal physics which has to be implemented in such a picture is to allow the
diquark to couple to two quarks. Such a coupling gives rise to quark exchange
between the diquark and the ``third'' quark within the baryon 
\cite{Cah89,Rei90}. In this first investigation we will restrict ourselves to
this minimal picture. Note that quark exchange is also required to reinstate
the Pauli principle within a Faddeev approach using effective diquarks.

A further important ingredient of our model is the effective
parametrization of confinement for the constituents of the baryon, the quarks
and diquarks. This is achieved by modifying the propagators of these
particles such that no Lehmann representation can be found for these
propagators. The price which has to be paid is an essential singularity at
timelike infinite momentum carried by these propagators\footnote
{Recent studies of a confining interaction within the 
Klein-Gordon equation revealed an essential singularity in the Isgur-Wise
function of heavy quarks \cite{Olso97}.}.
Note that such a
singularity prohibits the use of dispersion relations. Nevertheless, we
believe that the benefits of such a description, the absence of unphysical
thresholds, is worth the loss of applicability of  dispersion relations.

Given the fact that there exists numerous models for baryons, different types
of bag \cite{Has78} and soliton \cite{Sky61,Alk96} models as well as
non--relativistic\footnote{For a full solution of the nonrelativistic 
Faddeev equation see ref. \cite{Gloz97}. These authors, however, conclude:
``As a result  any nonrelativistic constituent quark model 
can at most be considered 
as a parametrization of the baryon energy levels, rather than as a dynamical
model for light three quarks systems. Certainly it will  not prove 
acceptable for future applications such as the description of 
electromagnetic form factors, hadronic decays, and other 
dynamical observables that are determined by the behaviour of
the baryon wave function and are generally much influenced by 
quark-quark potential parameters.''}  
\cite{Gel64,Kar68,Fai68} and 
relativistic potential models \cite{Fey71}, 
one is tempted to argue that there is no need for another
baryon model. A closer inspection, however, reveals that most, if not all, of
these models display severe shortcomings when applied to intermediate energy
reactions. Furthermore, at low energies these models describe \eg the nucleon
reasonably but not overwhelmingly well. These remark applies to models being
so different as for example the Skyrmion versus the MIT bag. Hybrid models
like the chiral bag \cite{Cho75,Rho94}  have deepened our insight but have
not been able to answer the question of the low--energy structure of the
nucleon conclusively. A very recent effort to understand quark correlations
in a solitonic background on the basis of the Nambu--Jona-Lasinio model
\cite{NJL61} have revealed that, at least within this model, quark binding
energies due to a self--consistent solitonic background and due to direct
two--quark and three--quark correlations are of the same order of magnitude
\cite{Zuc97}.  Therefore, one might argue that low--energy observables are
not sufficient to determine the structure of baryons even qualitatively. On
the other hand, high precision data in a momentum regime where quarks
certainly are dominant over collective (solitonic) effects but interactions
are still highly non--perturbative will help to clarify issues related to the
structure of baryons. The interpretation of these data, on the other hand,
requires a not too complicated modelling of the corresponding reactions. In
view of these remarks, we believe that there is still need for a baryon model
adapted to the intermediate energy regime.

This paper is organized as follows: In sect.\ \ref{Model} we describe our
model: Constituent quarks and diquarks interact via a Yukawa interaction
which represents the fact that a diquark may decay into two quarks. We
discuss a modification of the quark and diquark propagators which allow to
mimic confinement efficaciously. In sect.\ \ref{BSE} we discuss the
diquark--quark Bethe--Salpeter equation for the nucleon. Special emphasis is
hereby put on the decomposition in the Dirac algebra. The components of the
nucleon spinor are expanded  in hyperspherical harmonics. Numerical solutions
including scalar diquarks are presented.  In sect.\ \ref{NFF}
various nucleon form factors are calculated using Mandelstam's formalism. The
vertex functions of external currents to quarks and diquarks are constructed
from the corresponding Ward identities such that gauge invariance and
symmetries are respected at the constituent level.  In sect.\
\ref{Conclusions} we summarize the results obtained so far and give an
outlook.

\section{The Model} 
\label{Model}

As noted in the introduction the basic idea of our model is to parametrize
complicated and/or unknown  structures within baryons by means of two--quark
correlations in the overall antisymmetrized color--antitriplett channels. 
Furthermore, it is assumed that scalar ($0^+$) and axialvector ($1^+$) diquarks
are sufficient to describe baryons. As we are interested mainly in the nucleon
we will only consider two flavors. Due to the Pauli principle the scalar and
axialvector diquarks are then isoscalar and isovector, respectively. The
minimal physics which has to be implemented in such a picture is to allow the
diquark to couple to two quarks. Such a coupling gives rise to quark exchange
between the diquark and the ``third'' quark within the baryon. In this first
investigation we will restrict ourselves to this minimal picture. Note also
that quark exchange is required to reinstate the Pauli principle on the
three--quark level within a Faddeev approach using effective diquarks
\cite{Rei90}.

Additionally, we want to modify the kinetic terms of the fields in a way
that will allow for an effective parametrization of confinement. We will do
this on the Lagrangian level only symbolically, and discuss the specific form
of these terms in connection with the propagators used in the Bethe--Salpeter
equation.

The above description may be formalized with the help of the following
Lagrangian: 
\be 
\label{L1} 
\L & = &  \bar q_A(x) (i\gamma^\mu\partial_\mu-m_q)f(-\partial^2/m_q^2)
q_A(x)
 + \Delta^\dagger_A (x)(-\partial_\mu\partial^\mu - m_s^2)f(-\partial^2/m_s^2) 
   \Delta _A (x)
\nonumber \\ 
& &{}
 - \frac{1}{4}F^\dagger_{\mu\nu}(x)f(-\partial^2/m_a^2)F^{\mu\nu}(x) + 
\frac{1}{2}m_a^2\Delta^\dagger _{Aa\mu }(x)f(-\partial^2/m_a^2)
\Delta^\mu _{Aa} (x) 
\nonumber \\ 
& &{} + \frac {\epsilon^{ABC}}{\sqrt{2}} \left(
 g_s q_C^{T}(x) C i\gamma^5 \frac{t_{\mathcal A}}2 q_B(x) \Delta_A^{*}(x) 
+g_s^{*} \Delta_A(x)\bar q_B(x) i\gamma^5 C  \frac{t_{\mathcal A}}2 
\bar q_C^T(x)
 \right)
\nonumber \\ 
& &{} + \frac {\epsilon^{ABC}}{\sqrt{2}} \left(
 g_a q_C^{T}(x) C \frac{i\gamma^{\mu}}{\sqrt{2}} \frac{t_{\mathcal 
S}^a}2 q_B (x)
 \Delta^{*}_{Aa\mu}(x)
-g_a^* \Delta_{Aa\mu}(x)\bar{q}_B (x)\frac{t_{\mathcal S}^a}2 
 \frac{i\gamma^{\mu}}{\sqrt{2}} C \bar{q}_C^{T} (x)\right) .
\ee 
Here $q(x)$ describes the constituent quark with mass $m_q$. The scalar and
axialvector diquark field are called $\Delta$ and $\Delta^\mu$, respectively;
their masses are denoted by $m_s$ and $m_a$. 
The diquark field strength tensor is
given by $ F^{\mu\nu} = \partial^\mu\Delta^{\nu} - \partial^\nu\Delta^{\mu}
+ [\Delta^{\mu},\Delta^{\nu}]$ as the axialvector diquark field is due to its
isospin of non--abelian nature. Note, however, that we will not take into
account the related self--interactions of this field. The function $f$ stands
as reminder that the tree--level propagators of these fields will be modified
in order to describe their confinement. The Yukawa couplings
between quarks and diquarks are denoted by $g_s$ and $g_a$. Note that these
interactions are renormalizable. Nevertheless, we will eventually substitute
$g_s$ and $g_a$ by some additional momentum--dependent factors in order to
study the influence of the fact that realistic diquarks are certainly not
pointlike. 

Writing down the Yukawa interactions there are some ambiguities
concerning the phases of the interaction term. As we have chosen a hermitian
interaction term  we expect
the coupling constants to be real. The notation using $g_{s,a}$ and
$g_{s,a}^*$ is nevertheless chosen in order to allow for a more complete
discussion. 

In the Lagrangian (\ref{L1}) color indices are denoted by capital 
and isospin indices by small letters.
The color coupling via the $\epsilon$--tensor is determined by the assumed
color antitriplet nature of the diquark  allowing for a color singlet
interaction term. The matrix $t_{\mathcal A}=\tau^2$ is the antisymmetric 
generator of the
isospin group, the matrices $t_{\mathcal S}^a=\{\tau^3, 
{\vect{1}}, \tau^1\} =
\{ {\vect{\tau}} \tau^2 \}$ are the symmetric generators.  

As we will use the Bethe--Salpeter equation in Euclidean space we will
perform a Wick rotation thereby obtaining an Euclidean action from (\ref{L1}). 
We will do this by choosing a positive definite metric and hermitian Dirac 
matrices, $\{\gamma _\mu , \gamma _\nu \} = 2\delta _{\mu\nu}$ and 
$\gamma_\mu^\dagger =\gamma _\mu$.

In order to mimic confinement we will choose the function $f$ in the
Lagrangian (\ref{L1}) to be
\be
f(x) = 1- e^{-d(1+x)},
\label{conff}
\ee
\ie the (Euclidean) quark and diquark propagators are given by
\be
S(p) &=&  \frac{i{p  \!\!\! /}-m_{q}}{p^2+m_{q}^2}
\left( 1-e^{- d(p^2+m_{q}^2)/{m_{q}^2}} \right) ,
\label{S} \\
D_{AB}(p) &=&   -\frac{\delta^{AB}}{p^2+m_{s}^2}
\left( 1-e^{-d(p^2+m_{s}^2)/{m_{s}^2}} \right),
\label{Ds} \\
D_{AaBb}^{\mu\nu} (p) &=& -\frac{\delta^{AB}\delta^{ab}
(\delta^{\mu\nu} + p^{\mu}p^{\nu}/{m_{a}^2})}
{p^2+m_{a}^2}
\left( 1-e^{-d(p^2+m_{a}^2)/{m_{a}^2}} \right).
\label{Da} 
\ee
Obviously, $f$ modifies these propagators as compared to the ones
for free Dirac, Klein--Gordon and Proca particles. It is chosen
such that it removes the free particle pole at $p^2=-m^2_i$, $i=q,s,a$
While $d=1$ is the minimal way to mimic confinement, we will also 
study the cases $d > 1$ where the propagators are less modified
for spacelike momenta. 
The
propagators (\ref{S},\ref{Ds},\ref{Da}) are free of singularities for all 
finite $p^2$.
However, they possess an essential singularity at timelike infinite momenta,
$p^2=-\infty$. There exists no Lehmann representation for these
propagators: They describe confined particles. With respect to our calculations
the interesting property is simply the non-existence of thresholds related to
diquark and quark production.\footnote{Note that these form of the
propagators prohibits the use of dispersion relations. This is unavoidable
because unitarity is violated by construction.} 

The form of the propagators (\ref{S},\ref{Ds},\ref{Da}) is motivated from 
studies in the Munzcek--Nemirovsky model \cite{Mun83}.
In this model the
effective low--energy quark--quark interaction is modelled as
$\delta$--function in momentum space. These leads to a quark propagator
without poles on the real $p^2$ axis and a behaviour for  large negative $p^2$
similar to the one of (\ref{S})\footnote{Note that an infrared cutoff
in the NJL model has similar effects \cite{Ebe96}.}.   
Recently, it has been  shown that within
this model diquarks are also confined if one includes a non--trivial
quark--gluon vertex \cite{Ben96}. The diquark propagator does not have a pole
and is indeed similar to the form (\ref{Ds},\ref{Da})  whereas meson masses
as given by poles of the corresponding propagators stay almost unaltered as
compared to the model with a tree level quark--gluon vertex. Thus, these
investigations give us some confidence that an effective description of
confinement via the  parametrizations (\ref{S},\ref{Ds},\ref{Da}) is indeed
sensible. Nevertheless, for comparison we will also use in the
following free tree--level propagators, \ie $f\equiv 1$.

As the color structure of the propagators is trivial, color indices will be
suppressed in the following. In this paper we will assume isospin symmetry and
therefore also suppress isospin indices.

\newpage 
\section{Diquark--Quark Bethe--Salpeter Equation for the Nucleon} 
\label{BSE}

\begin{figure}
\begin{center}
\epsfig{file=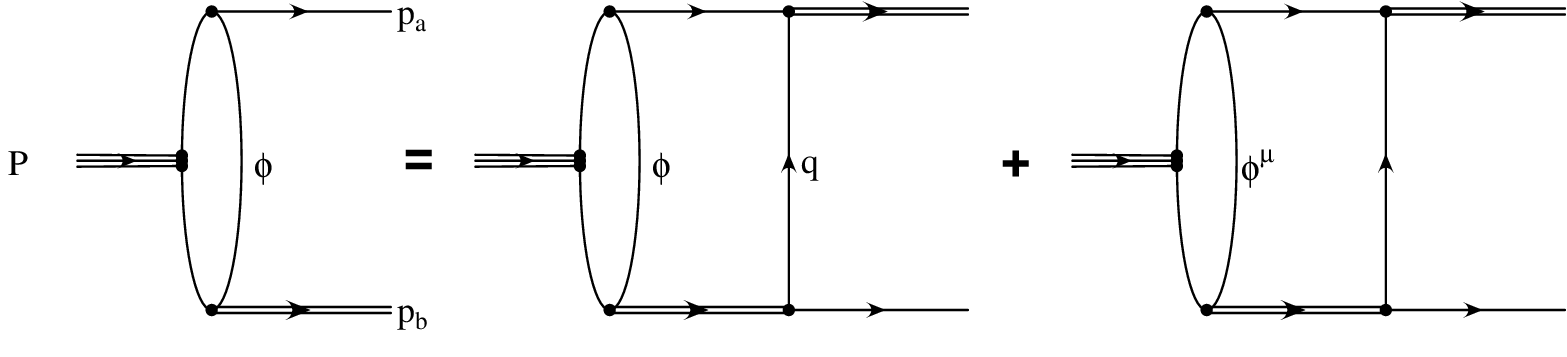,height=4cm,width=15cm}
\epsfig{file=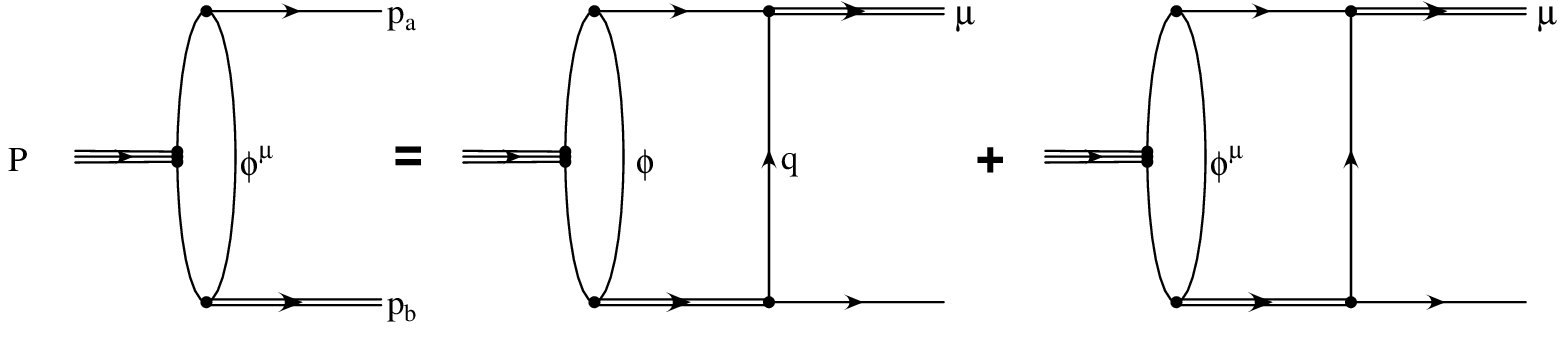,height=4cm,width=15cm}
\end{center}
\caption{\label{BSEfig} The Bethe--Salpeter equation for the scalar and axial
nucleon vertex functions.}
\vspace{0.3cm}
\end{figure}

Using the Lagrangian (\ref{L1}) the Bethe--Salpeter equation for the 
diquark--quark bound state  vertex functions is given by
\be
\Phi(p) &= & - |g_{s}|^{2} \int \frac{d^4p^{\prime}} {(2\pi)^4}
\gamma _5 S(-q) \gamma _5 S(p^{\prime}_{a}) D(p^{\prime}_{b})  \Phi(p^{\prime})
\nonumber \\
& &{}
+ \sqrt{\frac 3 2} g^*_{s} g_{a} \int \frac{d^4p^{\prime}}{(2\pi)^4} 
\gamma^{\mu}  S(-q) \gamma_{5} S(p^{\prime}_{a})
D^{\mu\nu}(p^{\prime}_{b}) \Phi^{\nu}(p^{\prime}) ,
\label{sc}
\\
\Phi^{\mu}(p)&=& -\sqrt{\frac 3 2} g_{s} g^*_{a}
\int\frac{d^4p^{\prime}}{(2\pi)^4} \gamma_{5} S(-q)  \gamma^{\mu} 
S(p^{\prime}_{a}) D(p^{\prime}_{b})  \Phi(p^{\prime})
\nonumber \\
& &{} -  |g_{a}|^{2}\int\frac{d^4p^{\prime}}{(2\pi)^4}
\gamma^{\lambda} S(-q) \gamma^{\mu} S(p^{\prime}_{a})
D^{\lambda\rho}(p^{\prime}_{b})  \Phi^{\rho}(p^{\prime}).
\label{ax}
\ee
These equations are diagrammatically represented in fig.\ \ref{BSEfig}. 
Hereby $P=p_a+p_b$ denotes the total momentum of the bound state and $p$ or
$p^{\prime}$ is the relative momentum between quark and diquark:
\be
p  =  (1-\eta) p_{a} - \eta p_{b}. \label{p}
\ee
The physical solution does not depend on the parameter $\eta \in [0,1]$,
however, as it is known from previous studies of Bethe--Salpeter equations the
accuracy of the numerical solution may depend on it (see \eg \cite{Jai93}). 
In these first exploratory
calculations we will only use the choice which renders the equations
(\ref{sc},\ref{ax}) as simple as possible, $\eta = 1/2$.

Using a static (momentum independent) approximation for the quark 
exchange, the Bethe-Salpeter equation (\ref{sc}, \ref{ax}) can be
reduced to a Dirac-type equation which can be solved almost 
analytically \cite{Buck}, see also \cite{Han95}. 
But the resulting baryon
vertex functions are then independent of the relative momentum between
quark and diquark and therefore not suitable for form factor 
calculations.

To avoid UV divergencies in the integral equations (\ref{sc}) and (\ref{ax})
it is convenient to work with a finite extension of the interaction 
(quark exchange)
in momentum space by modifying the exchanged quark propagator
according to
\be
S(q) \rightarrow S(q)\frac{\Lambda^2}{q^2 +\Lambda^2}.
\ee
While it was shown in ref. \cite{Kus97}, that this choice leads 
to sensible results for the nucleon amplitudes, 
there is certainly the need to calculate 
this ``form factor'', which can be related to the diquark Bethe-Salpeter
amplitude, from a microscopic diquark model. Work in this direction 
is under progress \cite{Hel97} (see also \cite{Ben96}).  

Note that the color algebra for the eqs.\ (\ref{sc},\ref{ax}) projected on
the the color singlet has simply provided an overall factor $-1$. The
corresponding factor for the color octet channel is $\frac 12$. As we assume
the color singlet channel interaction to be attractive the color octet one is
repulsive. Thus, using confined quarks and diquarks, no states in the 
octet channel can be generated.

The isospin Clebsch--Gordan coefficients  are given by
\be
\pmatrix{1&-\sqrt{3}\cr -\sqrt{3}&-1 \cr}
\ee
and have been already absorbed in the coefficients in eqs.\
(\ref{sc},\ref{ax}). 

Eqs. (\ref{sc},\ref{ax}) determine the Bethe--Salpeter vertex functions $\Phi
(P,p)$ and $\Phi^\mu (P,p)$. They are related to the Bethe--Salpeter wave
function by multiplication with the free two--particle propagator,
\be
\Psi (P,p) & = & S(P/2+p) D(P/2-p) \Phi (P,p), 
\label{Psidef} \\
\Psi^\mu (P,p) & = & S(P/2+p) D^{\mu\nu} (P/2-p) \Phi^\mu (P,p).
\label{Psimudef} 
\ee

\subsection{Decomposition in Dirac Algebra}

As we want to describe a nucleon with positive energy we use a corresponding
projection from the very beginning \ie we require that the Bethe--Salpeter
vertex functions and wave functions are proportional to a positive energy Dirac
spinor for a particle with momentum $P$ and helicity $s$, $u(P,s)$. For later
convenience we define the following Dirac--matrix--valued quantities
\cite{Mey94}:
\be
\chi(P,p) &=& \sum_{s=\pm 1/2} \Phi(P,p) \otimes \bar u(P,s), \quad
\chi^\mu(P,p) = \sum_{s=\pm 1/2} \Phi^\mu(P,p) \otimes \bar u(P,s),
\label{chidef} \\
\omega(P,p) &=& \sum_{s=\pm 1/2} \Psi(P,p) \otimes \bar u(P,s), \quad
\omega^\mu(P,p) = \sum_{s=\pm 1/2} \Psi^\mu(P,p) \otimes \bar u(P,s).
\label{omegadef} 
\ee
These are then proportional to the positive energy projector
\be
\Lambda^+ = \sum_{s=\pm 1/2} u(P,s) \otimes \bar u(P,s) =
\frac{1}{2}(\vect{1}+  \hat P\!\!\! /  )
\ee
where $\hat P^\mu = P^\mu/iM$ is the unit momentum vector parallel to the bound
state momentum. Using the projector property $(\Lambda^+)^2= \Lambda^+$ the
vertex functions can be written as
\be
\Phi(P,p) & = & \chi (P,p) \Lambda^+ u(P,s), \\
\Phi^{\mu}(P,p) & = & \chi^{\mu}(P,p) \Lambda^+ u(P,s).
\ee

The most general Dirac decomposition for 
the scalar function $\chi (P,p)$ involves four independent Lorentz scalar 
functions, \eg
\be
\chi(P,p)  =  a_1(P,p) + a_2(P,p){P\!\!\! /} + a_3(P,p){p\!\!\! /} +
a_4(P,p)\frac{1}{2} ({P\!\!\! /}{p\!\!\! /} - {p\!\!\! /}{P\!\!\! /}),
\ee
whereas the one for $\chi^{\mu}(P,p)$ decomposes into twelve Lorentz 
axialvector functions, \eg
\be
\hspace{-0.6cm}
\gamma _5 \chi^{\mu}(P,p)& = & b_1(P,p) P^{\mu} + b_2(P,p) p^{\mu} +
b_3(P,p) \gamma^{\mu} + b_{4}(P,p)P^{\mu} {P\!\!\! /} +
  b_{5}(P,p)p^{\mu}{p\!\!\! /} 
% +b_{6}(P,p) P^{\mu}{p\!\!\! /} 
+ \ldots . 
\ee
Projection onto positive energies leads to two and six independent components,
respectively. It is advantageous to choose the Dirac matrices in these
expansions to be eigenfunctions to $\Lambda^+$. In addition, the expansion
coefficients are required to fulfill the Dirac equation for a particle of mass
$\pm iM$. This leads to
\be
\chi (P,p) &=& S_1(P,p)\Lambda^+ + S_2(P,p) \Xi \Lambda^+ 
\label{Sdef} \\
\nonumber \\
\chi^\mu (P,p) &=&  A_1(P,p)\gamma_5\hat P^\mu \Xi \Lambda^+ 
+  A_2(P,p)  \gamma_5 \hat P^\mu \Lambda^+
\nonumber \\
&+&  B_1(P,p)\gamma_5\hat p^\mu_T \Xi \Lambda^+ 
 +   B_2(P,p)\gamma_5\hat p^\mu_T \Lambda^+
\nonumber \\
&+& C_1(P,p)  \gamma _5 (i(\hat P^\mu \vect{1} -\gamma^\mu)
			 - \hat p^\mu_T \Xi ) \Lambda^+ 
  + C_2(P,p) \gamma _5 (i(\hat P^\mu \vect{1} +\gamma^\mu) \Xi 
			  -\hat p^\mu_T)\Lambda^+
\nonumber \\
\label{Adef}
\ee 
where
\be
\Xi = -i( p\!\!\! /  -  \hat P \cdot  p ) /\sqrt{p^\mu_Tp^\mu_T}  
\label{xi}
\ee
is a Dirac matrix whose interpretation is obvious in the center-of-mass frame
of the bound state, see below, and
\be
p^\mu_T = p^\mu - \hat P^\mu  (\hat P \cdot p) \quad {\rm and} \quad
\hat p^\mu_T = p^\mu_T /\sqrt{p^\mu_Tp^\mu_T}
\ee
are projections of the relative momentum on the direction transverse to the
total momentum, $ \hat p^\mu_T P^\mu =0$. 
Note that the normalization applied in eq. (\ref{xi}) leads to 
easily interpretable expressions of the amplitudes in the rest frame.
Nevertheless, for numerical treatment, we will modify the normalization 
slightly, see below.

The notation in eqs. (\ref{Sdef}), (\ref{Adef}) 
is such that an index 1 refers to an operator in the
expansion being a solution to a Dirac equation with mass $+iM$ (upper
component in the rest frame of the bound state) and an index 2 refers to an
operator in the expansion being a solution to a Dirac equation with mass $-iM$
(lower component in the rest frame of the bound state).

Let us detail this for the case of the scalar diquark. Obviously, in the rest
frame of the bound state, $P=(0,0,0,iM)$, the operators read
\be
\Lambda^+ = \pmatrix{1&0\cr 0&0\cr}, \quad 
\Xi\Lambda^+ =\pmatrix{0&0\cr\vect{{\hat p}}\vect{\sigma}&0\cr}.
\ee
Thus, multipliying the amplitude $\chi$ with an positive energy spinor $u$,
using the definition of $\chi$ (\ref{chidef}) and the relation
$\bar u(P,s) u(P,t) = \frac E M \delta _{st} = \delta _{st}$ (the
last equal sign is only valid because we have specialized to the rest frame
of the bound state)   one obtains:
\be
\Phi (P,p) = \pmatrix{\vect{1}  S_1(P,p) \cr
\vect{{\hat p}}\vect{\sigma} S_2(P,p)\cr}.
\label{PHI}
\ee
As anticipated this is the spinor for a ${\frac 1 2}^+$ particle. The more
complicated case of the vector spinor $\Phi^\mu$ relating to the amplitude
with the axialvector diquark as a constituent is given in Appendix A.

\subsection{Expansion in Hyperspherical Coordinates}

In order to obtain a numerical solution of the diquark--quark Bethe--Salpeter
equation we expand the amplitudes $S_i, A_i, B_i$ and $C_i$, $i=1,2$, in
hyperspherical harmonics. The basic idea  hereby is the would--be O(4)
symmetry of the Bethe--Salpeter equation if the exchanged particle would be
massless. Thus, the hope is that only a few orders are sufficient for a
numerically precise solution which will be indeed the case, see \cite{Kus97}
and the next section.

First, let us recall that in four dimensional spherical coordinates,
$$
p^\mu = p (\cos \phi \sin \theta \sin \psi , \sin \phi \sin \theta \sin
\psi , \cos \theta \sin \psi , \cos \psi ),
$$ 
the spherical spinor harmonics are given by
\be
{\mathcal Z}_{njlm} (\psi , \theta , \phi ) = 
\sqrt{ \frac {2^{2l+1}(n+1)(n-l)!(l!)^2}{\pi(n+l+1)!}}
\sin ^l \psi C^{1+l}_{n-l} (\cos \psi ) {\mathcal Y}^j_{lm} (\theta , \phi ) .
\ee
Hereby the $C^{1+l}_{n-l} (\cos \psi )$ are the Gegenbauer polynomials
\cite{AS65}. The ones with O(3) angular momentum $l=0$ are identical to the
Chebychev polynomials of the second kind,
\be
C^{1}_{n} (\cos \psi ) = T_n (\cos \psi ) = \frac {\sin (n+1) \psi }{\sin
\psi },
\ee
obeying the orthogonality relation
\be
\int_{0}^{\pi}\!\! d \psi \sin^2 \psi\:  T_n(\cos \psi) T_m(\cos \psi) =
\frac{\pi}{2} \delta_{n m} 
\label{ortho}
\ee
The O(3) spherical spinor harmonics ${\mathcal Y}^j_{lm} (\theta , \phi )$ are
the usual ones familiar from relativistic quantum mechanics. They obey the 
relation
\be
{\mathcal Y}^j_{j\pm 1/2,m} (\hat p) = - \vect {{\hat p}} \vect{\sigma} 
{\mathcal Y}^j_{j\mp 1/2,m} (\hat p) .
\ee
Using the usual spherical harmonics $Y_{lm}$ the explicit expression is given 
by 
\be
{\mathcal Y}^j_{lm} (\theta , \phi ) = \frac 1 {\sqrt{2l+1}}
\pmatrix{\pm \sqrt{l\pm m + \frac 1 2} Y_{l,m-1/2} \cr
             \sqrt{l\mp m + \frac 1 2} Y_{l,m+1/2} \cr}
\ee
for $j=l+\frac 1 2$ and $j=l-\frac 1 2$, $l>0$, respectively.

The amplitudes in eqs. (\ref{Sdef}), (\ref{Adef})
$S_i, A_i, B_i$ and $C_i$, $i=1,2$, are expanded in Chebychev
polynomials, e.g. 
\be
S_i(P,p) = \sum_{n=0}^{\infty} i^n S_{in} (P^2,p^2) T_n (\hat P \cdot \hat p) 
\label{GGBexp}
\ee
and analogous for the functions $A_i,B_i$ and $C_i$.  
 
\subsection{Numerical method}
For the numerical solution of the Bethe-Salpeter equation we will 
restrict ourselves in the following to $0^+$ diquarks, i.e 
the solution of eq. (\ref{sc}) with $g_a=0$.
The inclusion of $1^+$ diquarks and accordingly the axialvector
amplitudes of the nucleon is much more involved and will be treated in a 
forthcoming publication. 
Furthermore we choose equal masses for the quark and the diquark, 
$m_q \equiv m_s$. Then all dimensional quantities in the Bethe-Salpeter 
equation can be expressed in units of the constituent mass $m_q$. 
To allow a comparison of our results with 
the ones reported in 
\cite{Kus97} we also 
choose, if not stated otherwise, $\Lambda=2 m_q$, which basically 
fixes the width of the interaction in momentum space. 

When working in the rest frame of the bound state we 
parametrize   
the relative momenta in eq. (\ref{sc}) according to 
\be
p'_{\mu} &=& p' (\cos \phi ' \sin \theta ' \sin \psi ' , \sin \phi ' 
\sin \theta '
 \sin\psi ' , \cos \theta ' \sin \psi ' , \cos \psi ' ), \\
p_\mu &=& p (0,0, \sin \psi , \cos \psi ).
\ee
By changing the normalization for the Dirac matrix in (\ref{xi}) to
\be
\tilde \Xi = -i( p\!\!\! /  -  \hat P \cdot  p ) /\sqrt{p^\mu p^\mu}  
\ee
and denoting the spatial part of the  normalized $4-$vector $p$  by
$\vect{{\tilde p}}$,
the generic structure of the integral equation can be written 
as
\be
\pmatrix{\vect{1}  S_1(P,p) & 0 \cr
\sin \psi \sigma_3  S_2(P,p) & 0 \cr}&=&-g_s^2 \mintpp 
K(p,p',P)
\pmatrix{\vect{1}  S_1(P,p') & 0 \cr
\vect{{\tilde p'}}\vect{\sigma} S_2(P,p') & 0 \cr}, \\
\nonumber\\*
K(p,p',P)&\sim&\gamma _5 S(-q) \gamma _5 S(p^{\prime}_{a}) D(p^{\prime}_{b})
\label{BSEPRI}
\ee
Using the expansion of the amplitudes in terms of
Gegenbauer polynomials (\ref{GGBexp}) and 
applying the orthogonality relation (\ref{ortho}), we
extract the expansion functions (amplitudes)
$S_{i,m}(p), \,(i=1,2; m=0.. \infty)$
and obtain after integration over $\phi'$
\be
\pmatrix{S_{1 m}(p) \cr
         S_{2 m}(p) \cr}
= g_s^2\sum_{n=0}^{\infty} i^{n-m}
%\int_{0}^{\pi}\!\!d \psi \sin^2 \psi 
%\int_{0}^{\pi}\!\!d \psi ' \sin^2 \psi ' 
%\int_{0}^{\pi}\!\!d \Theta ' \sin \Theta ' 
\int \!\! d \Sigma  
\int_{0}^{\infty} \!\! d p' p'
\pmatrix {\tilde K_{11} &  \tilde K_{12}\cr
          \tilde K_{21} &  \tilde K_{22}\cr} \cdot 
\pmatrix{S_{1 n}(p') \cr
         S_{2 n}(p') \cr}.
\label{BSEfinal}
\ee
By taking the spin degeneracy into account we reduced the 
problem to two coupled integral equations, 
where 
\mbox{$\tilde K_{i,j}, (i,j=1,2)$}
now denote real scalar quantities and can be calculated straightforwardly.
The remaining angular integrals are given by
\be
\int \!\! d \Sigma= 
\int_{0}^{\pi}\!\!d \psi \sin^2 \!\psi 
\int_{0}^{\pi}\!\!d \psi ' \sin^2 \!\psi ' 
\int_{0}^{\pi}\!\!d \Theta ' \sin \Theta '.  
\ee
For the actual solution the Gegenbauer expansion has to be terminated 
at a finite $n=m_{max}$  and due to the structure of eq. (\ref{BSEfinal})
$m$ is in the range  $0 \leq m \leq m_{max}$.
Equation (\ref{BSEfinal}) is now the  
most convenient expression of the Bethe-Salpeter equation 
for numerical treatment.
While the $\Theta '$ integration 
can be done analytically, we perform the $\psi$ and $\psi '$ integration
numerically by means of Gaussian quadratures with appropriate
weights \mbox{($w(z)=\sqrt{1-z^2}, z=\cos \psi$)}.
The final step is then to discretize $p$ and $p'$ with $k_{max}$ 
grid points to generate a $k_{max} \times k_{max}$ mesh in momentum space.
Actually we map the interval $[0,\infty]$ to $[-1,1]$ and determine 
the grid points according to a Gaussian quadrature with Gauss-Legendre 
weights. In this way a 
$(2\!\cdot\!m_{max}\!\cdot\!k_{max} \times 2\!\cdot\!m_{max}\!\cdot\!k_{max})$
matrix is generated which can be  treated as an eigenvalue problem; for
a given bound state mass $M$ (which appears nonlinearly in the integral
equation) we seek the corresponding coupling constant $g_s$. 
In order to test the numerics we solved the eigenvalue problem
by straightforward diagonalization as well as by an iteration method 
and found, within numerical accuracy, identical results. When applying 
the iteration method one starts with initial guesses for the 
amplitudes and iterates the matrix equation until $g_s$ converges.   

As a further check of our numerical procedure
we also 
solved the Bethe-Salpeter equation  involving explicitly the wavefunction
$\Psi(P,p)$ 
by iteration,
% where $z=\hat{P}\cdot\hat{p}$, and
%correspondingly $z'=\hat{P}\cdot\hat{p'}$:
%
\begin{eqnarray}
 \label{SBSE}
 \Phi(P,p)& =& 
       - g_s^2\int \frac{d^4 p'}{(2\pi)^4} H(p,p') 
\Psi(P,p')\\
    \Psi(P,p)& =& D(P/2-p)S(P/2+p) \Phi(P,p).
\end{eqnarray}
Note that the corresponding  kernel $H\sim\gamma_5 S(-q) \gamma_5$ is 
independent of 
$P$, or $M$, which in principle allows a fast determination of $M$ with
given $g_s$. First we use
an adequate decomposition for $\Psi(P,p)$ as in
eq. (\ref{PHI}),
\be
   \Psi(P,p)= {\vect{1}G_1(P,p) \choose \sin \psi\, \sigma_3 G_2(P,p)}.
\ee
We then also  
expand the free two-particle propagator 
$D(P/2-p)S(P/2+p)$ in terms of Gegenbauer
polynomials (obtained in the case of tree-level propagators analytically, 
in the case
of confining propagators numerically).
This defines the coefficients $u_{ij}$ via the relation
{\small
\be
D(P,p)S(P,p){\vect{1}S_1(P,p) \choose \sigma_3 \sin \psi\, S_2(P,p) }= 
 \sum_{k_1=0}^{\infty} i^{k_1} \pmatrix{u_{11}^{k_1} & u_{12}^{k_1} \cr
  \sigma_3 \sin \psi \: u_{21}^{k_1} & \sigma_3 \sin \psi \: u_{22}^{k_1}\cr} 
\times \nonumber\\*  
T_{k_1}(\cos \psi) {S_{1}(P,p) \choose S_{2}(P,p)}.
\ee
}
The final step is the
projection onto the expansion coefficients $S_{im}(p)$ and $G_{im}(p), i=1,2,$
to yield
\be \label{S1}
 {S_{1m}(p) \choose S_{2m}(p)}  = 
   g_s^2 \sum_{n=0}^{n_{max}} i^{n-m} \int \!\! d \Sigma
\int_0^{\infty} d|p'|\,|p'|^3 
  \pmatrix {\tilde H'_{11} &  \tilde H'_{12}\cr
          \tilde H'_{21} &  \tilde H'_{22}\cr} \cdot
     {G_{1n}(p') \choose G_{2n}(p)} 
\ee 
\be \label{G1}
 {G_{1n}(p) \choose G_{2n}(p)}=\sum_{m=0}^{m_{max}} \sum_{k_1=0}^{m+n_{max}} 
  i^{k_1+m-n} \pmatrix{u_{11}^{k_1} & u_{12}^{k_1} \cr
       u_{21}^{k_1} & u_{22}^{k_1}\cr} \cdot {S_{1m}(p) \choose S_{2m}(p)}
\ee
\hspace*{7cm} {\footnotesize \parbox[t]{5cm}{$|k_1-m| \leq
  n \leq  k_1+m \\
  k_1+m \equiv n$ mod 2,}}

which is suitable for an iteration procedure.
Note, that 
we used the following addition theorem for Gegenbauer polynomials:
\be  
 T_n(z) T_m(z) = T_{|n-m|}(z) + T_{|n-m|+2}(z) + \dots + T_{n+m}(z).
\ee

Once the kernel $H(p,p')$ was computed 
on a momentum mesh we iterated $\Phi(P,p)$ (expanded up to
order $m_{max}$) taking into account $\Psi(P,p)$ 
up to order $n_{max}$, as expressed
in (\ref{S1}). 

In the limit $n_{max} \rightarrow \infty$ both methods are by construction
equivalent. In the
actual calculations $n_{max}=m_{max}+3$ proved to be sufficient as the 
eigenvalues for a given $m_{max}$ quickly converged due to the asymptotic
behaviour of the kernel for higher orders $m$ and $n$.

Before discussing our numerical results we want to mention that 
the linear Bethe-Salpeter 
equation provides no a priori normalization condition.
While a physical normalization of the bound state amplitudes will be 
discussed in the next section here we simply  demand $S_{10}(p_1)=1$
($p_1$ denotes the first grid point of the momentum mesh), which 
provides a preliminary overall normalization.  

\subsection{Numerical results}
\label{results}

\begin{figure}[t]
\centerline{{
\epsfxsize 8.2cm
\epsfbox{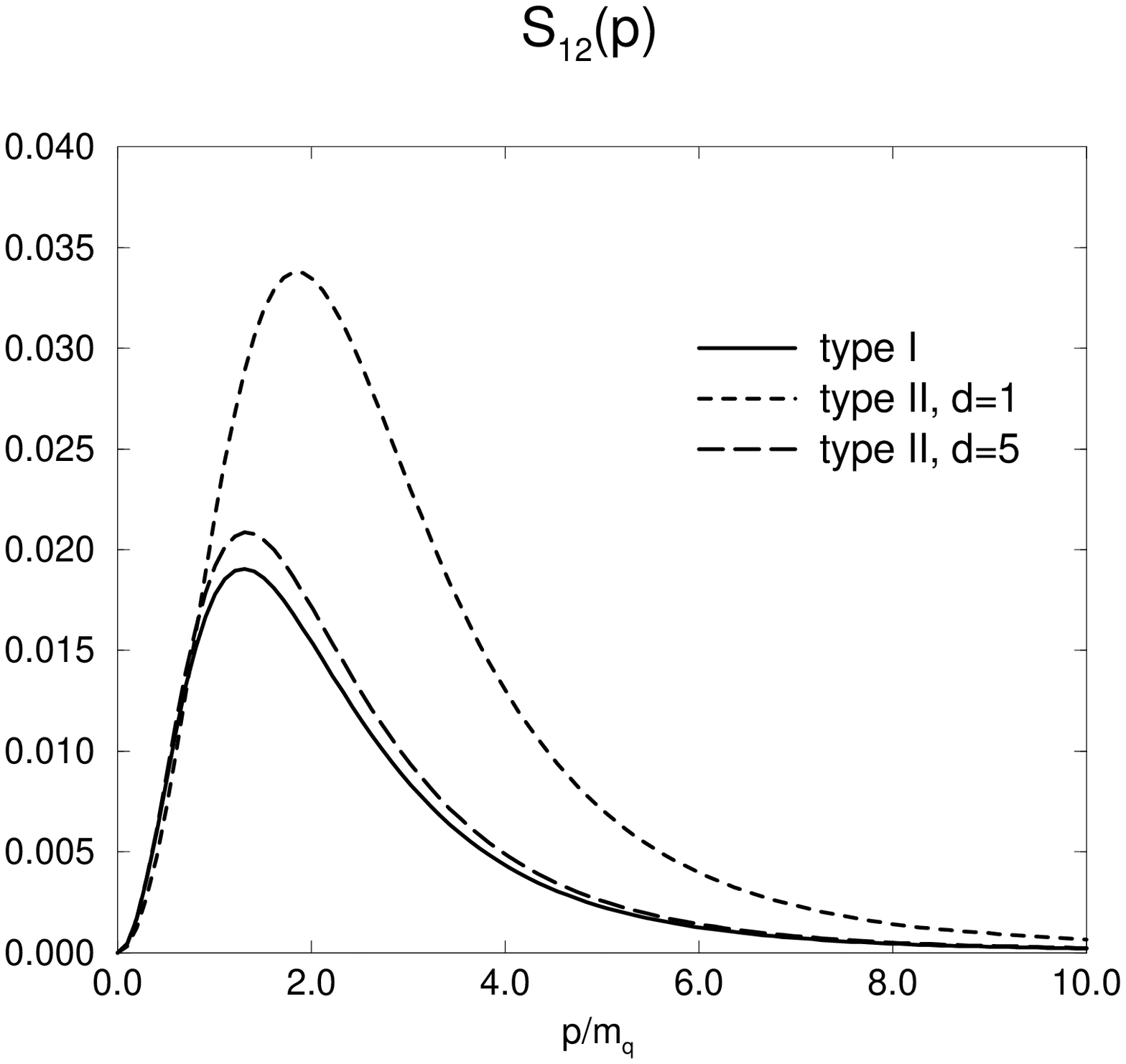}}{
\epsfxsize 8.2cm
\epsfbox{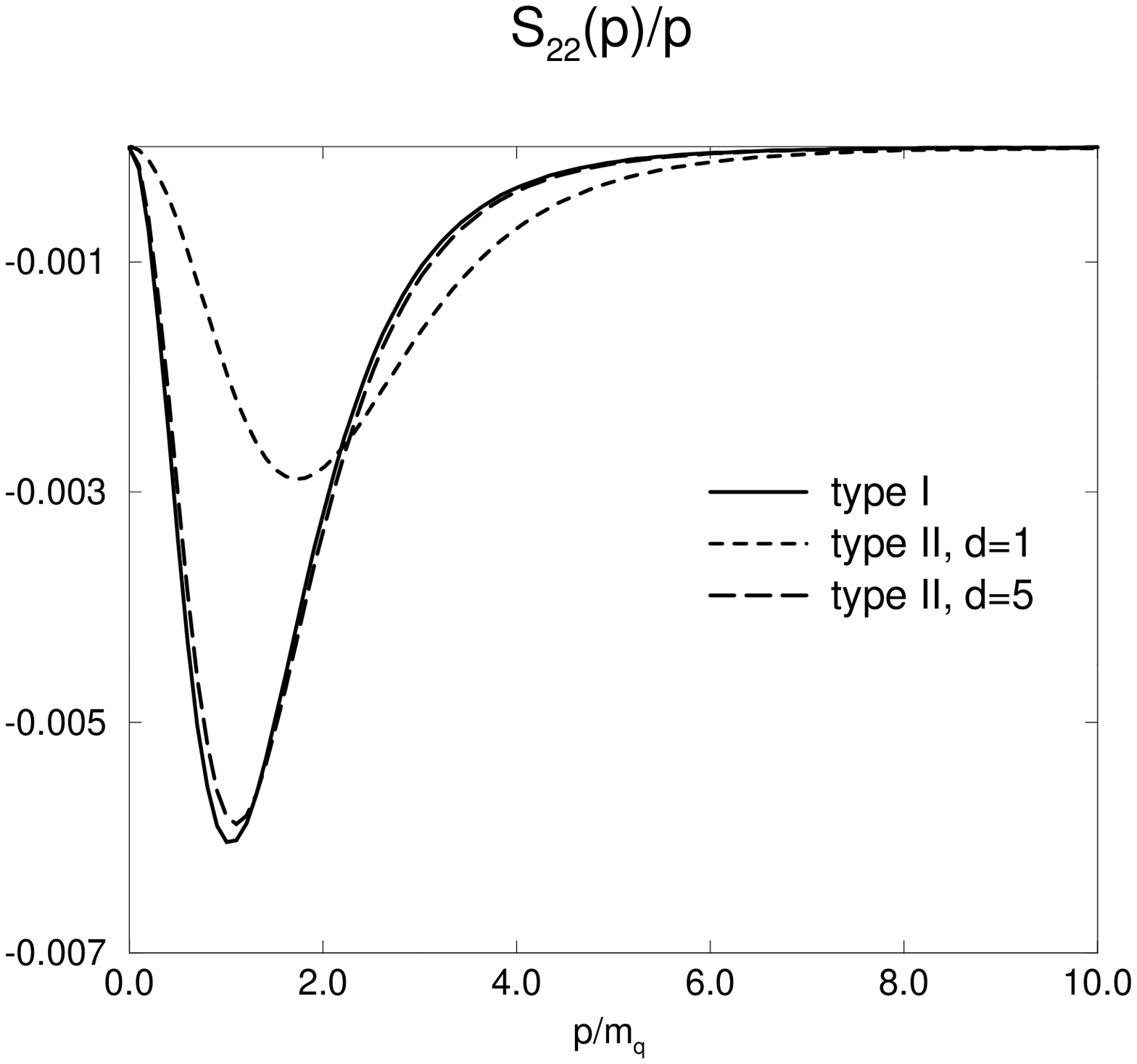}}}
\caption{Bethe-Salpeter amplitudes $S_{10}(p)$...$S_{22}(p)$
calculated with type I and type II propagators  
(with different values for $d$) and with $k_{max}=30$.\label{allamp}}  
\end{figure}

In this subsection we discuss  our numerical results.
By examining the convergence properties of the integral 
equation we observed that for $k_{max} \geq 30$ the eigenvalues
and eigenvectors do not depend any more on the size of the momentum mesh.  
The numerically obtained amplitudes are shown in 
figure \ref{allamp}.
Type I denotes the calculations where quark and diquark propagators 
with poles are included (\cf $f\equiv 1$ in eq. (\ref{conff}));
type II denotes calculations with 
confining propagators. When doing the calculation with propagators
of type II we also vary $d$, the damping factor in the exponentials.
As it is seen in the figures if  $d$ is increased,
the amplitudes approach the results obtained with type I propagators.
For the case $d=10$ there is no visible deviation from 
the type I amplitudes.
This feature is expected since for a large damping factor basically 
only the timelike properties of 
the quark and diquark 
propagators 
are  affected by the exponential. 
We note that the amplitudes for the type I calculations are in 
agreement with the results in \cite{Kus97}, which provides a
check of our calculation.

\noindent
\begin{table}[h]
\centering
\begin{tabular}{||c||c|c|c||}
\hline
    \multicolumn{4}{||c||}{\rm type I} \\
\hline
  $m_{max}$ &  M $= m_q $ & M $=1.9 m_q$ & M $=1.99 m_q$\\ 
\hline
  0 & 15.8164  & 9.9237   & 8.2132   \\ 
  1 & 16.1141  & 9.8853   & 8.1729   \\ 
  2 & 16.0809  & 9.8488   & 8.1496   \\ 
  3 & 16.0812  & 9.8485   & 8.1494   \\ 
\hline
\end{tabular}
\caption
{Eigenvalues of the Bethe-Salpeter equation with type I 
propagators for different 
nucleon masses when the expansion in terms of Gegenbauer polynomials 
is terminated at $m=m_{max}$.\label{typI}}
\end{table}

From figure (\ref{allamp}) 
it can now be seen that the employed expansion in terms
of Gegenbauer polynomials converge rapidly: The absolute value of 
subsequent 
amplitudes (compare \eg $S_{11}$ with $S_{10}$) decreases roughly 
by one 
order of magnitude. So higher orders $(n \geq 3)$ can be safely 
neglected.

\newpage
\begin{figure}[t]
\centerline{{
\epsfxsize 8.2cm
\epsfbox{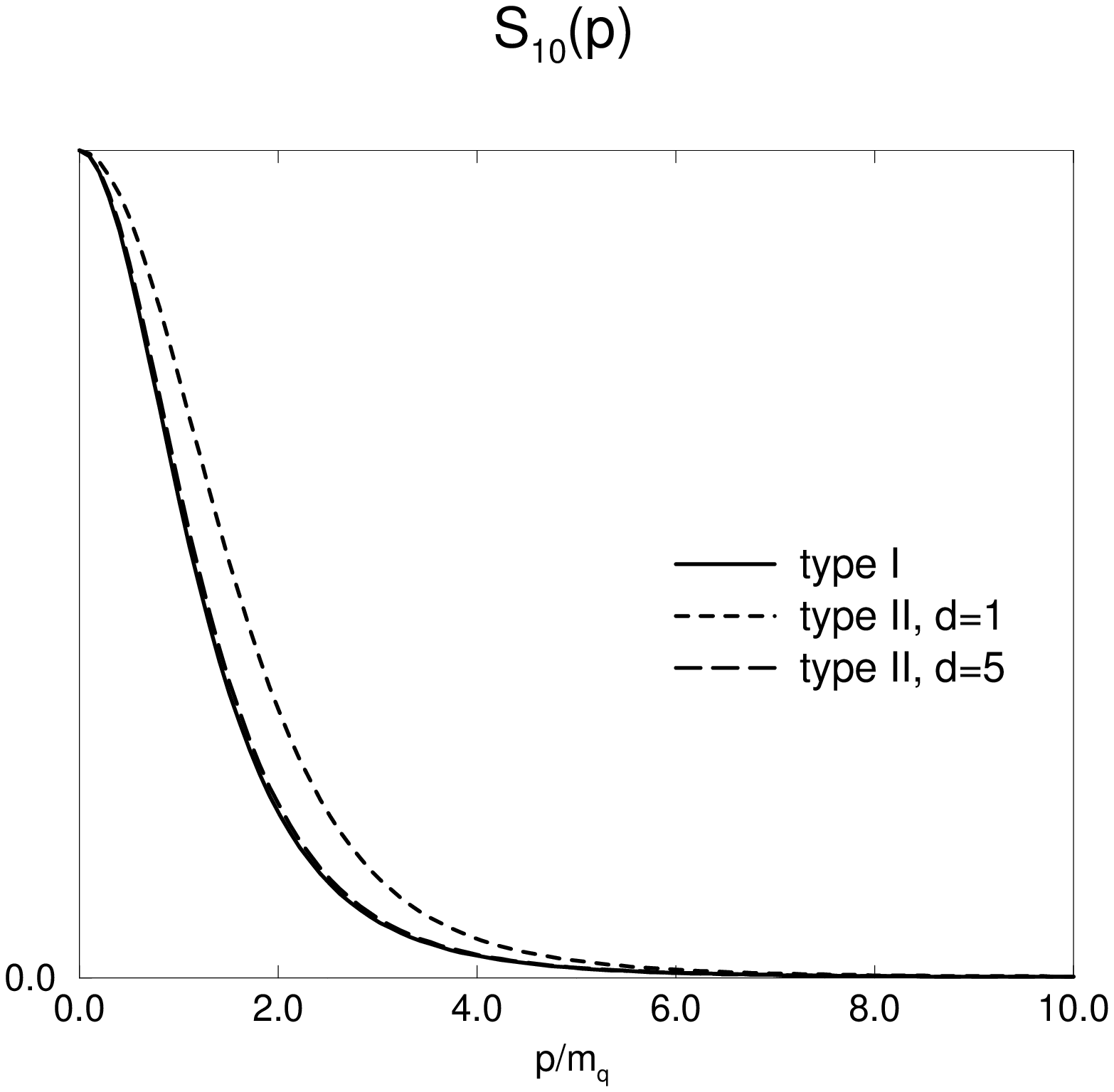}}{
\epsfxsize 8.2cm
\epsfbox{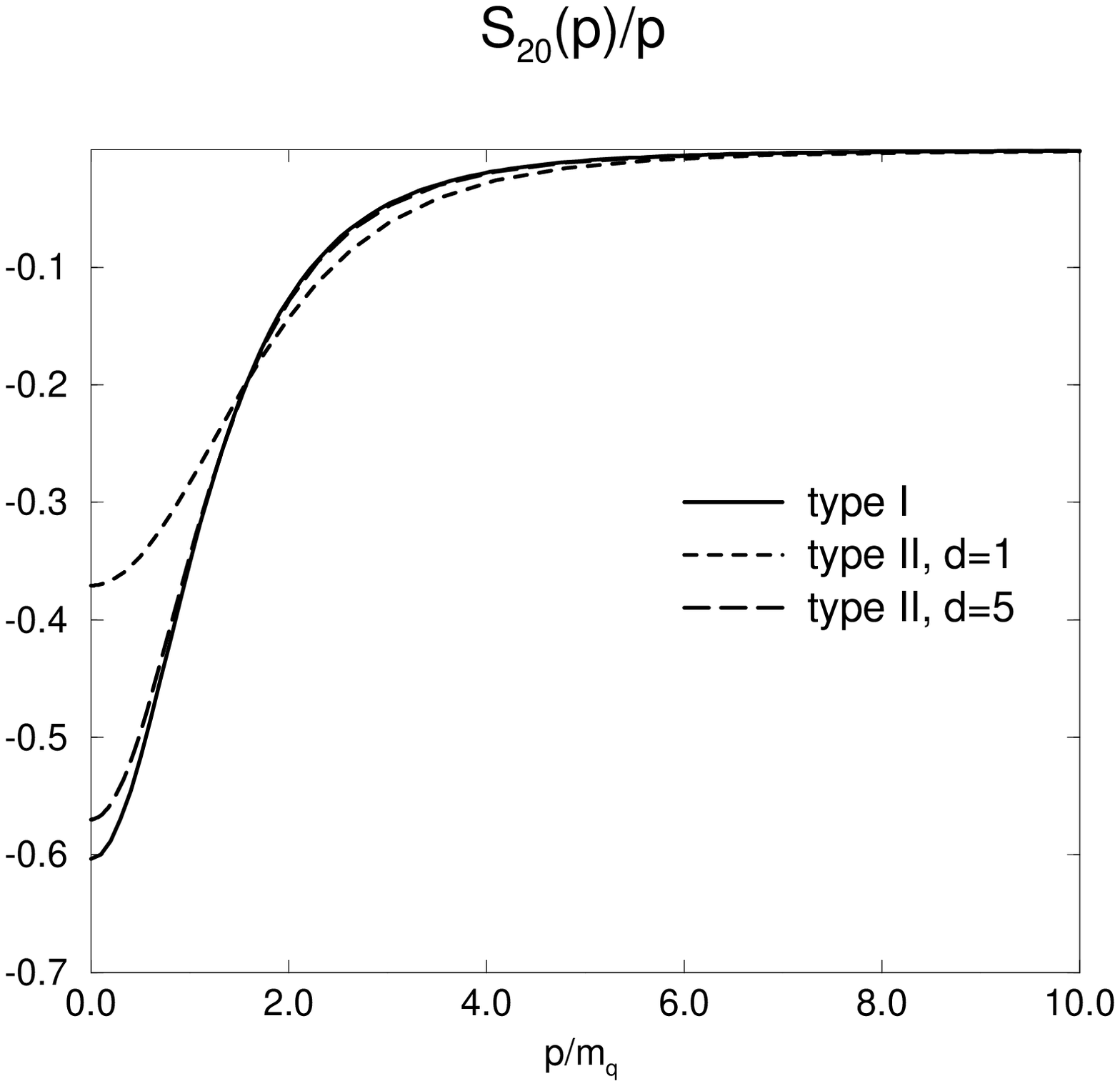}}}
\end{figure}
\begin{figure}[t]
\centerline{{
\epsfxsize 8.2cm
\epsfbox{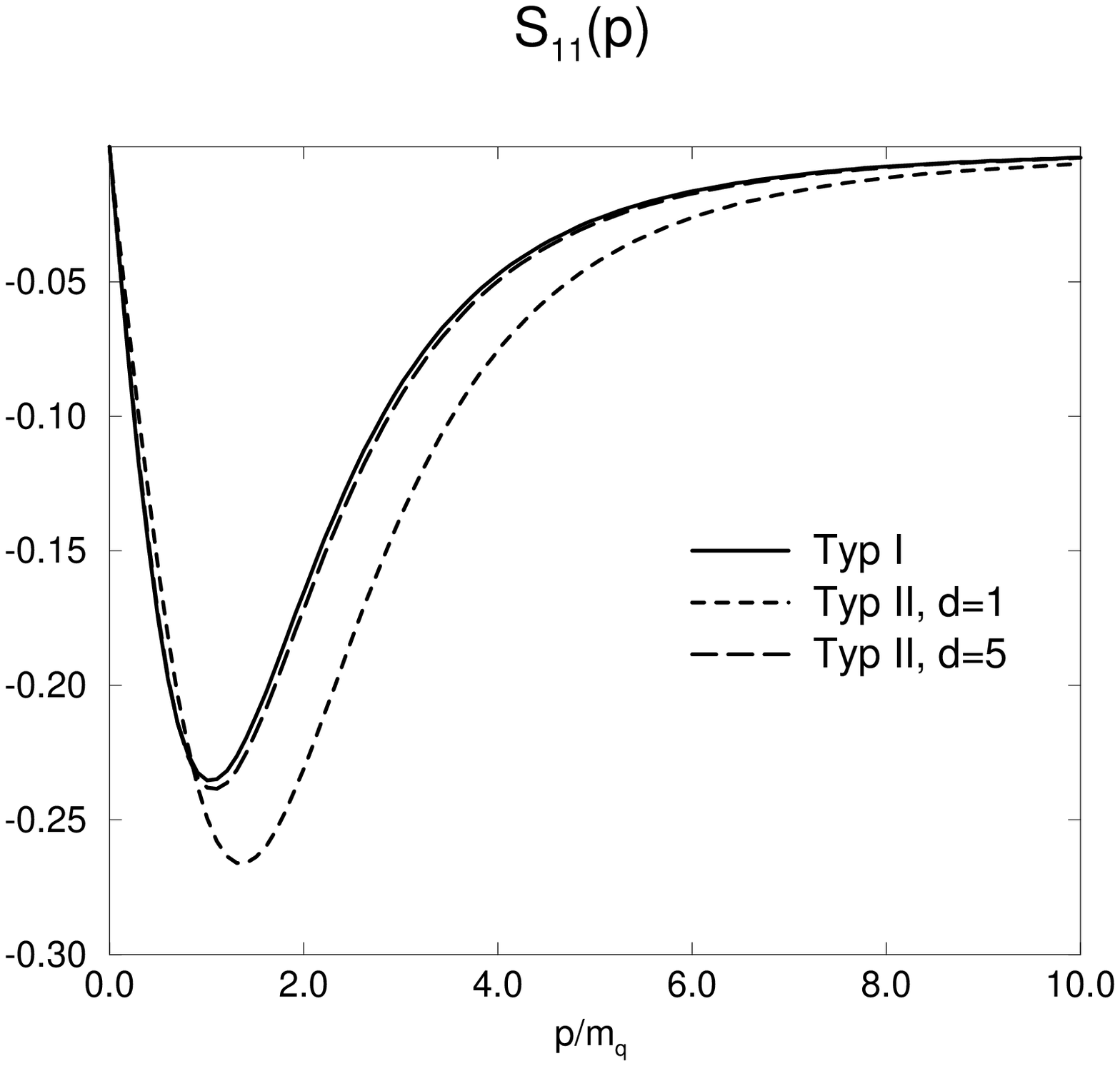}}{
\epsfxsize 8.2cm
\epsfbox{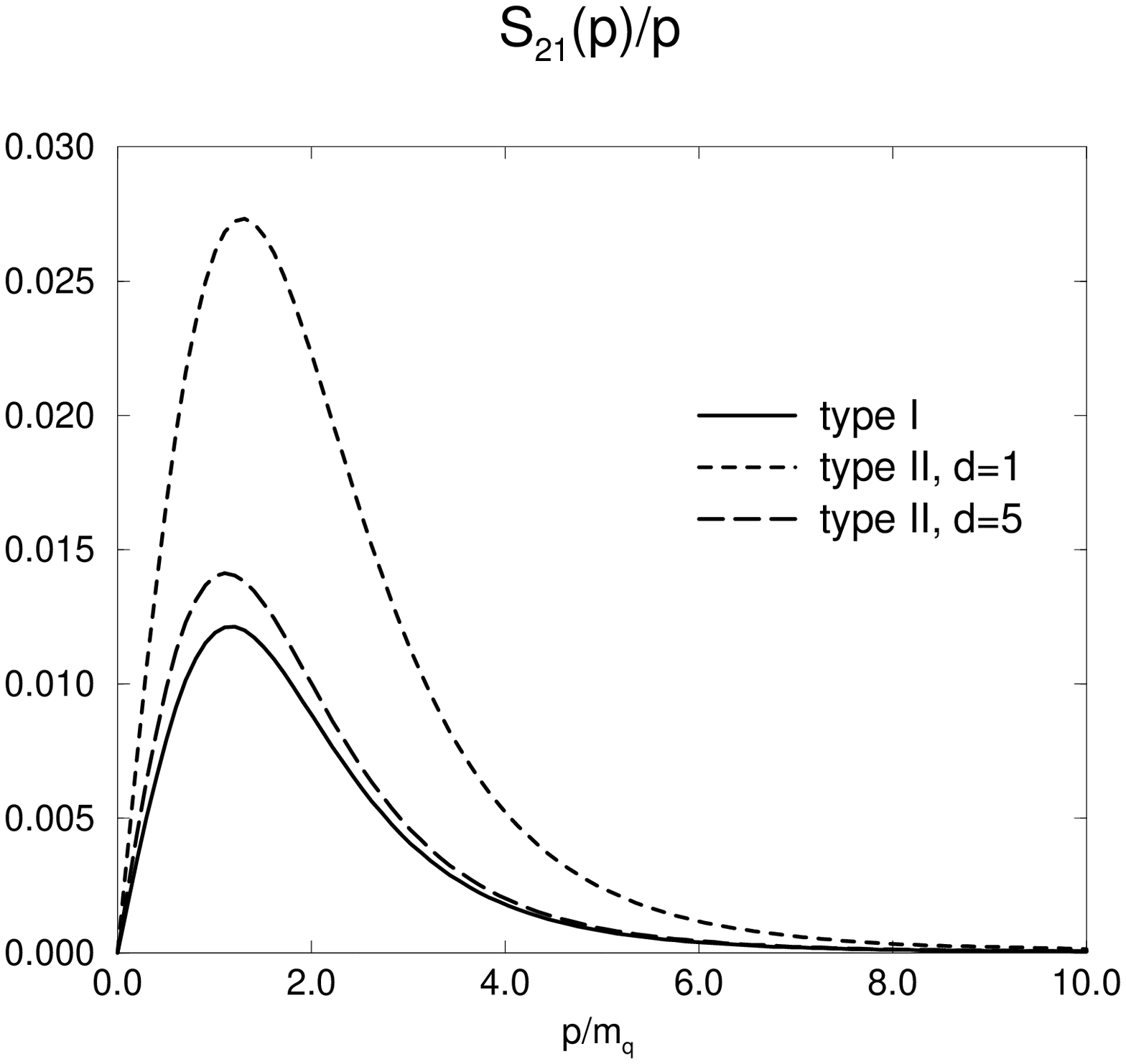}}}
\end{figure}
\newpage

\noindent
\begin{table}[h]
\centering
\begin{tabular}{||c||c|c|c||}
\hline
    \multicolumn{4}{||c||}{\rm type II, d=1} \\
\hline
  $m_{max}$ &  M $= m_q$  & M $=1.9  m_q$ & M $=1.99 m_q$\\ 
\hline
  0 & 17.7627 & 14.0058 &  13.5747   \\ 
  1 & 18.2856 & 14.0938 &  13.6373  \\ 
  2 & 18.2502 & 14.0485 &  13.5916  \\ 
  3 & 18.2507 & 14.0487 &  13.5918  \\ 
\hline
\end{tabular}
\caption
{Eigenvalues of the Bethe-Salpeter equation with type II and $d=1$
propagators for different 
bound state masses when the expansion in terms of Gegenbauer polynomials 
is terminated at $m=m_{max}$.\label{typII}}
\end{table}

In order to stress this fact, we 
show in table (\ref{typI}) and (\ref{typII})
how the eigenvalues 
of the integral equation converge when more orders of  the 
Gegenbauer expansion
are included in the calculation, \ie $m_{max}$ is increased.
The same behaviour can be observed when 
the eigenvectors (amplitudes) are examined; graphically they then
become undistinguishable. 
Thus, we conclude that especially 
in the ``physical region'' of the binding energy ($m_q \leq M \leq 1.9 m_q$) 
sufficient
convergence is achieved with $n_{\rm max}=2$, thereby including $3$ orders 
in the expansion.

Though the two methods to solve the Bethe-Salpeter equation
(one for the nucleon vertex function and one for the wave function)
mentioned in the last section are expected to give equivalent results, 
deviations might arise in the numerical integration procedure.
The kernel
$K(p,p';P)$, including the free two-particle propagator
approaches zero much faster than $H(p,p')$ 
for large values of $p'$ and
for higher orders $m,n$ in the Gegenbauer expansion.
Therefore, using the second method, the number of grid points 
has to be increased to obtain the same numerical
accuracy; we typically
chose $50$ points for the momentum mesh.
Then the results agree very well within numerical accuracy.

\begin{figure}[]
\centerline{{
\epsfxsize 8.2cm
\epsfbox{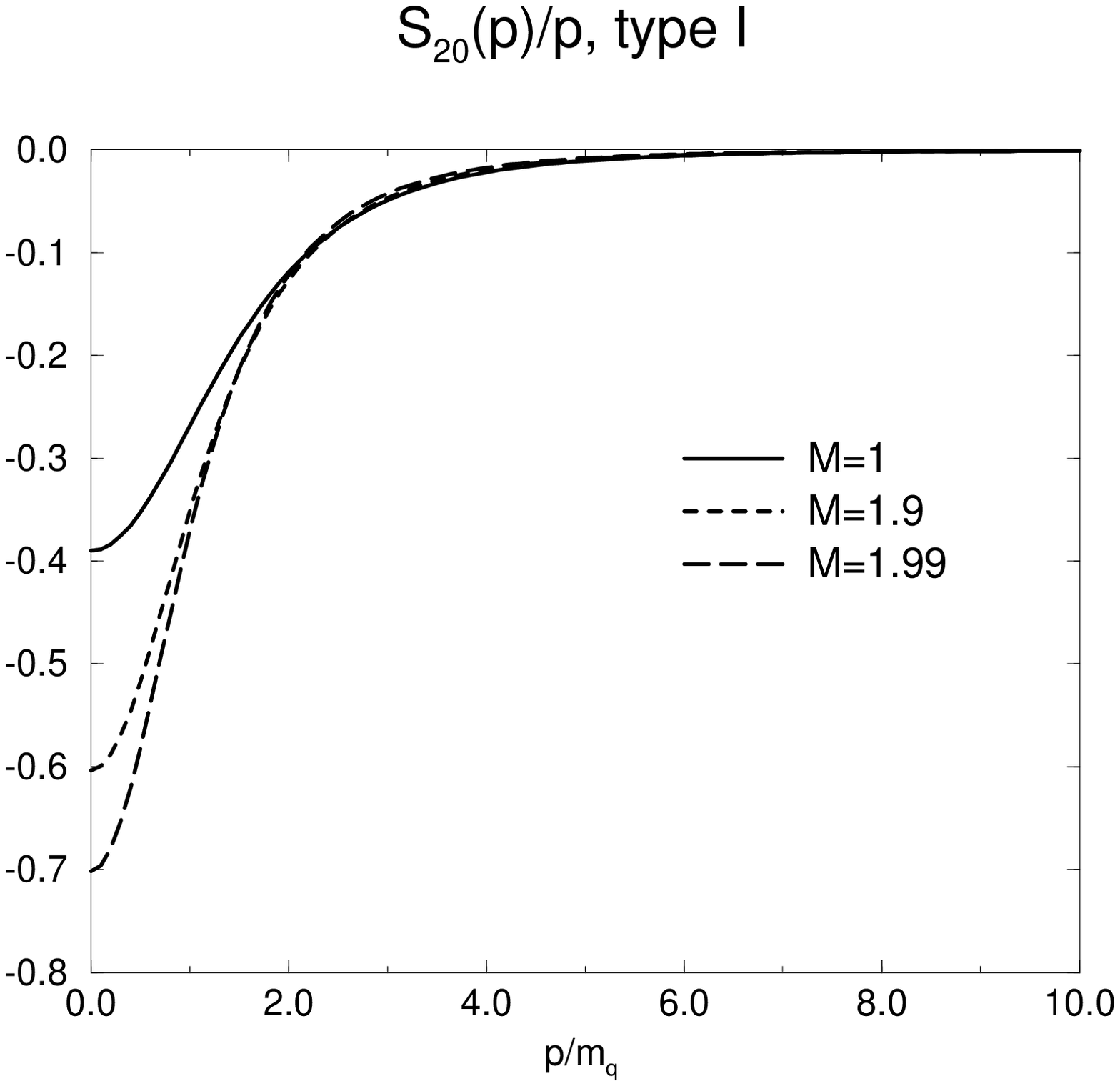}}{
\epsfxsize 8.2cm
\epsfbox{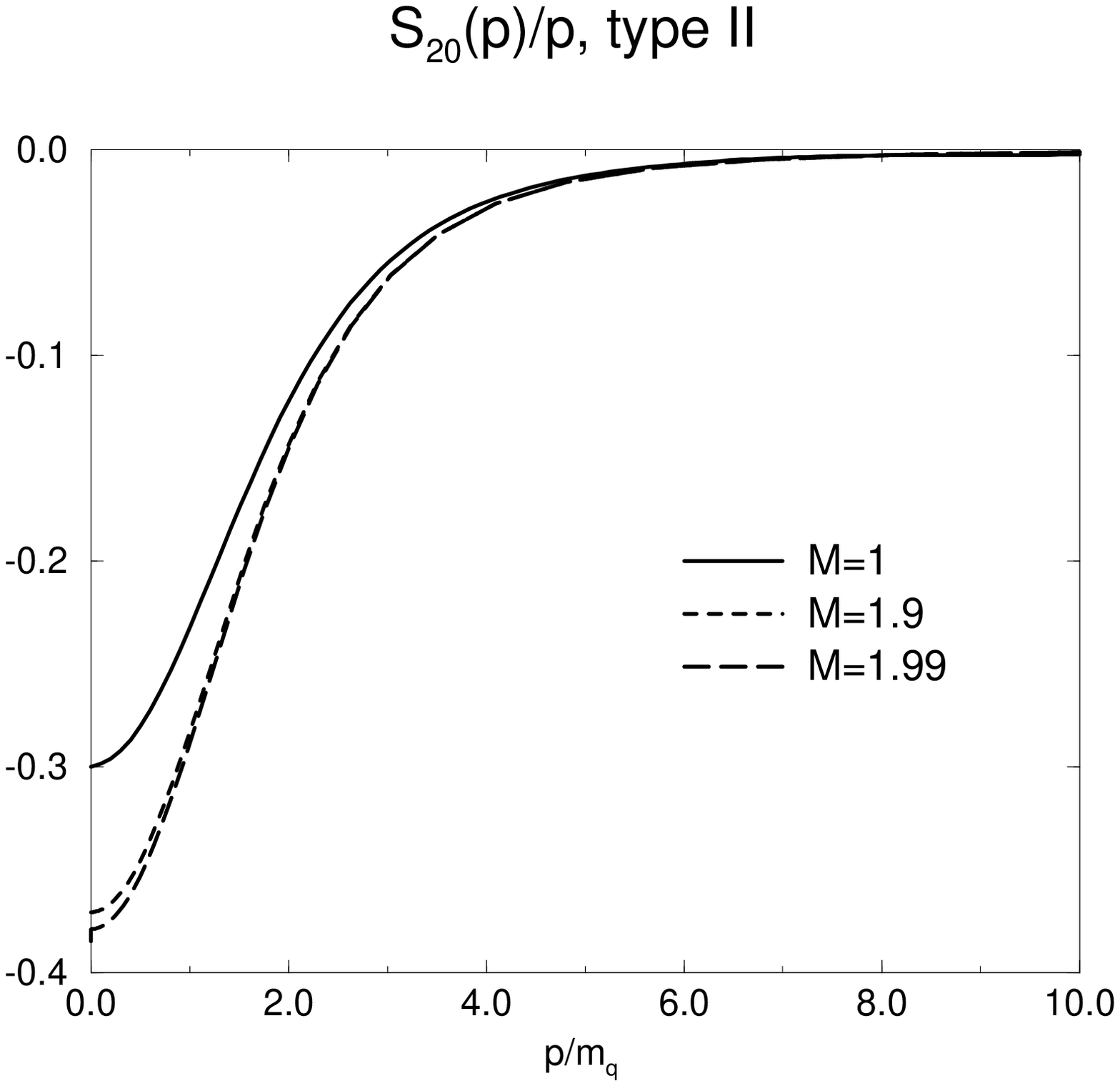}}}
\caption{The lower component $S_{20}(p)$ for $M=m_q, 1.9 m_q, 1.99 m_q$
using type I propagators, left hand figure, and type II propagators 
with $d=1$, right hand figure.\label{s2}}
\end{figure}

Having in mind that $S_{2}(p)$ describes the lower component
of the nucleon Dirac field (which is usually neglected, \eg in 
nonrelativistic calculations), it is surprising that we obtain rather 
large ``small'' components. This was also observed in ref. 
\cite{Kus97}. To study this in greater detail we plot 
in figure (\ref{s2}) the amplitude 
$S_{20}(p)$ for different values of the bound state mass. 
For both types of propagators the maximum of the amplitudes $S_{20}(p)$
increases when the binding energy $E_B=2m_q-M$ tends to $0$. 
Note however that this result depends on the chosen representation
of the fermion field (in our case the Dirac 
representation)\footnote{We thank F.\ Lenz for this remark.}. 
We will investigate 
in the next section whether the large lower
component
of the nucleon spinor has effects on observables 
where a $\gamma_5$ is involved in the coupling to the 
external current ($g_A(Q^2)$ and $g_{\pi NN}(Q^2))$
because this $\gamma$-matrix induces a mixing between 
upper and lower components 
which give rise to  contributions to these observables.
\\
\begin{figure}[]
\vspace{0.5cm}
\centerline{{
\epsfxsize 8.5cm
\epsfbox{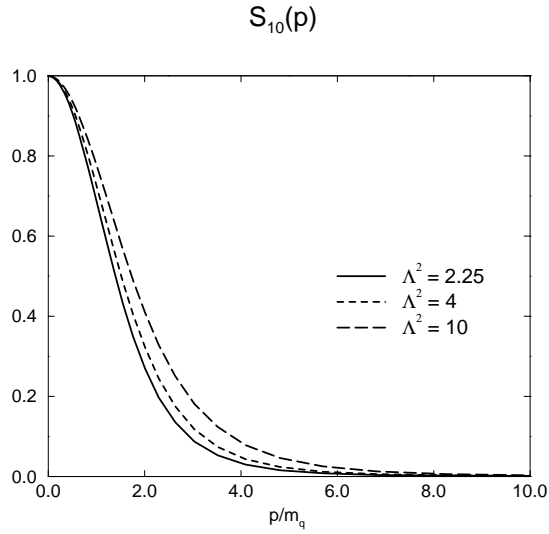}}}
\caption{$S_{10}(p)$ is plotted for type II and $d=1$ 
propagators for different 
values of $\Lambda^2$ (in units of $m_q^2$) 
which controls the range of the 
interaction.\label{width}} 
\end{figure}

The dependence of the amplitudes on the width of the interaction
can be seen from fig. (\ref{width}). Although we restricted ourselves in the
plot to a calculation of $S_{10}(p)$ for type II propagators with $d=1$
and a bound state
mass of $M=1.9 m_q$ the observed behaviour is quite generic: When $\Lambda$
is increased, higher modes of the interaction are included,
leading  to a broader  amplitude  in momentum space. 
Correspondingly,  in coordinate space 
the interaction become shorter ranged  leading to a narrower amplitude.
Although this behaviour  would allow  to fit $\Lambda$ e.g. to the
mean square radius of the nucleon, we will not do this in our 
following calculation.
 
\begin{figure}[t]
\centerline{{
\epsfxsize 8.5cm
\epsfbox{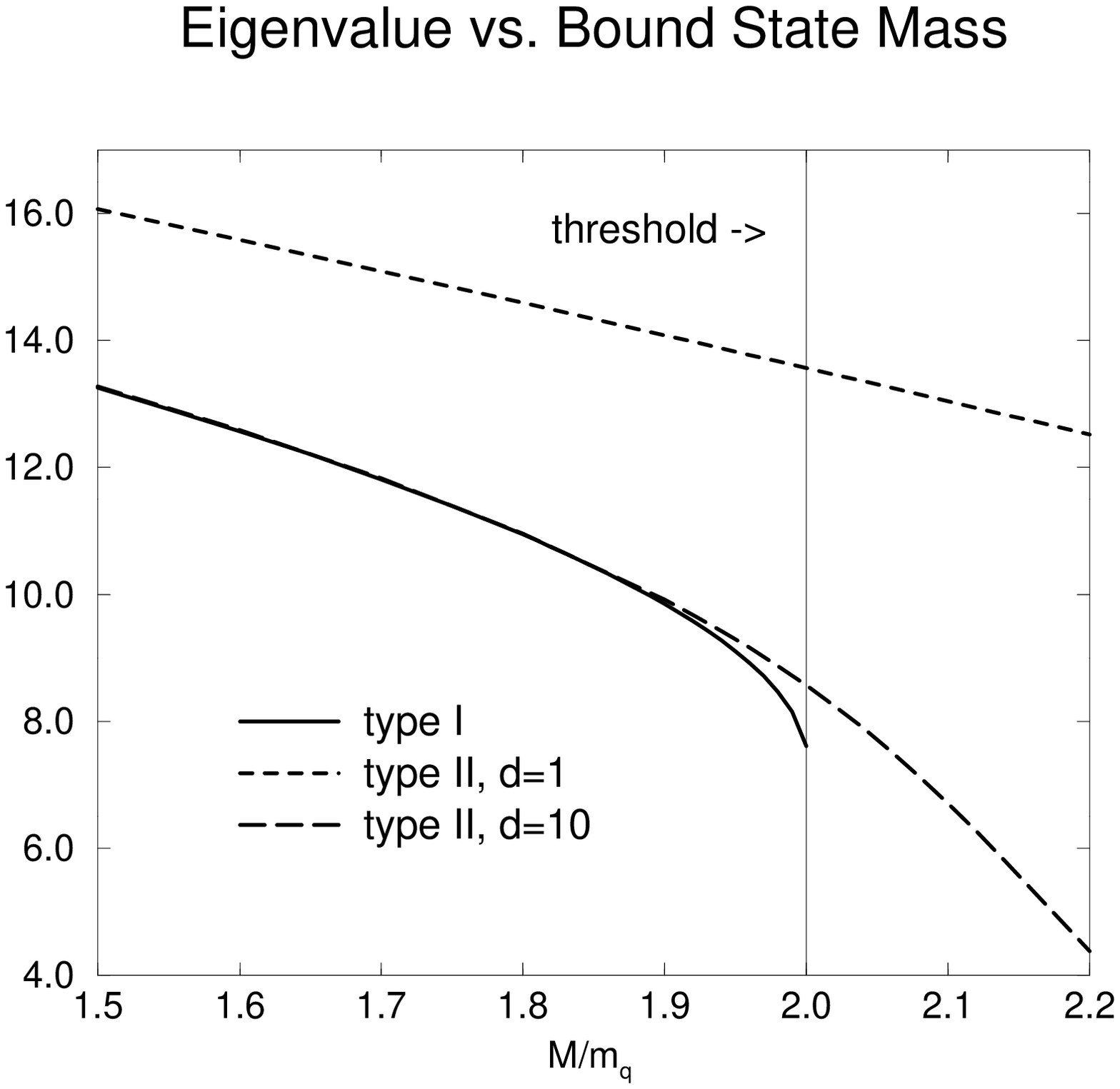}}}
\caption{Eigenvalues of the integral equation as functions of the 
bound state mass $M$. In case of type I propagators, the influence of 
the quark-diquark threshold is clearly visible.\label{eigen}} 
\end{figure}

In order to demonstrate the main advantage of our approach, namely 
working not only with free (type I) propagators but also with confining 
propagators (type II), we display  in figure \ref{eigen}
the calculated coupling constant $g_s$ ($\sim$ eigenvalue) as a function
of the nucleon  mass.
For type I propagators the eigenvalues near threshold 
start to decrease rapidly. Ideally they should display an infinite slope
for
$M=2m_q$, because there  the bound state decays into its constituents.
To observe this feature a much more refined numerical method (suitable
for extremely  loose bound states) would be necessary. Our numerical
method, which is very robust in a wide range of bound state 
masses cannot account for this property, 
but signals 
the nearby threshold by a strong sensitivity of $g_s$ on $M$.
When using now the confining propagators we get a totally different 
behaviour: The eigenvalues are not affected by the ``threshold''
any more
(which in this case is no threshold at all). The function
$g_s$ is a smooth function and nothing signals the possibility 
of the decay of the bound state. In this way confinement is realized in our
approach: By choosing confining propagators for the constituents of the
nucleon, we exclude its unphysical decay into free quarks and diquarks.
Even when the damping factor $d$ is increased, so that for $M \leq 1.9$ 
the eigenvalues and eigenvectors  
of a type I calculation are basically recovered, 
near ''threshold''
we again observe confinement.

%\vspace{1.2cm}
Before we describe in the next section the application of 
the so far obtained 
amplitudes in a calculation of baryonic matrix elements,
we want to mention 
that the numerical solutions of the Bethe-Salpeter equation
can be parametrized for further applications very effectively.
By examining the infrared and ultraviolet asymptotics of the amplitudes 
we find simple rational functions with parameters fitted to the 
numerical solutions. For details we refer to appendix \ref{fit}.

\section{Nucleon Form Factors} 
\label{NFF}
In order to test the validity of the proposed quark-diquark picture of
the nucleon, it is not enough just to solve the corresponding Bethe-Salpeter
equation, since its solution, the Bethe-Salpeter amplitudes 
(or the vertex functions) have no physical interpretation by themselves.
It is therefore necessary to use the solution of the Bethe-Salpeter
equation  
to calculate physical observables as, \eg form factors.
Especially we are interested in electromagnetic form factors of proton and 
neutron, the pion-nucleon form factor $g_{\pi NN}$ 
and the axial form factor $g_A$. 
These observables are accessible  
by coupling an appropriate  probing external current to the nucleon.
\\
In ref. \cite{Ish95,Asa95} a comparable calculation to our approach 
was performed in the context 
of the NJL-model (with unconfined quarks and diquarks), but only 
static properties (vanishing momentum transfer) of the nucleon
were considered; in ref. \cite{Hel95} spacelike form factors were
calulated (also within the NJL model) 
but assuming a static quark exchange between  quark and 
diquark and therefore working with a ``trivial'' nucleon vertex function.
A calculation within the Salpeter approach, instead of the here employed 
fully relativistic Bethe-Salpter equation, is reported 
in ref. \cite{Kei96a} and \cite{Kei96b}. 
So far, there exist no 
work in which a) the BS equation was solved without any approximation
for the quark exchange and b) the resulting solution was used 
to evaluate observables for finite momentum transfer to the nucleon.
See, however, ref. \cite{Kus97} for a calculation of nucleon structure 
functions within this approach.
\\
At this point we make the following remarks: The so far developed 
picture of a nucleon consisting of a quark and a scalar diquark 
with equal masses is certainly to na\"\i ve for a correct descripton
of the nucleon and therefore quantitatively agreement with experimental
results cannot be expected. Nevertheless we believe that the 
reported studies are a necessary step in the development 
of a complete baryon model. Therefore we not even try 
to fit e.g. $\Lambda$ or $d$ to experimental numbers. Our aim  is 
rather to observe, whether certain nucleon properties 
can be described reasonably. 
\\
The calculation of hadronic matrix elements between bound state 
vertex functions
is conveniently performed in Mandelstam's formalism \cite{Man55}, 
which will be
introduced in the next subsection.

\subsection{Mandelstam's Formalism}
In order to calculate the matrix element of an external current 
operator (generically denoted by $\hat O$)
between bound state amplitudes we use Mandelstam's formalism \cite{Man55}
(see also \cite{Asa95}),
which reads in momentum space  
\be
\langle \hat O_{(\mu)} \rangle
 &=& \langle N(P_f,S_f)|\hat O_{(\mu)} |N(P_i,S_i) \rangle
 = \frac{1}{\sqrt{4 E_{P_f} E_{P_i}}} J_{(\mu)}(P_f, P_i), 
\\
J_{(\mu)}(P_f,P_i) 
 &=& \mintpf \mintpi
{\bf \bar \Psi} (P_f, p_f)\Gamma_{\hat O}(p_f,P_f;p_i,P_i) 
{\bf \Psi} (P_i,p_i).
\label{MANDELSTAM}
\ee
Here $P_i, p_i$ and $P_f=P_i+Q$, $p_f$ 
denote the total and the relative
momenta of the quark-diquark bound state before and after the 
interaction with the external current, while   
${\bf \Psi}$ and ${\bf \bar \Psi}$ 
are  the properly normalized Bethe-Salpeter 
wave functions. By writing explicitly the quark and diquark 
propagators, which are included in the definition of the BS-wave function,
we obtain Mandelstam's prescription  in a representation, which 
contains the nucleon vertex function, calculated in the previous section.
\begin{figure}
%\begin{center}
%
%\epsfig{file=quark3.eps,height=4cm,width=15cm}
%\epsfig{file=diquark3.eps,height=4cm,width=15cm}
\centerline{{
\epsfxsize 8.2cm 
\epsfbox{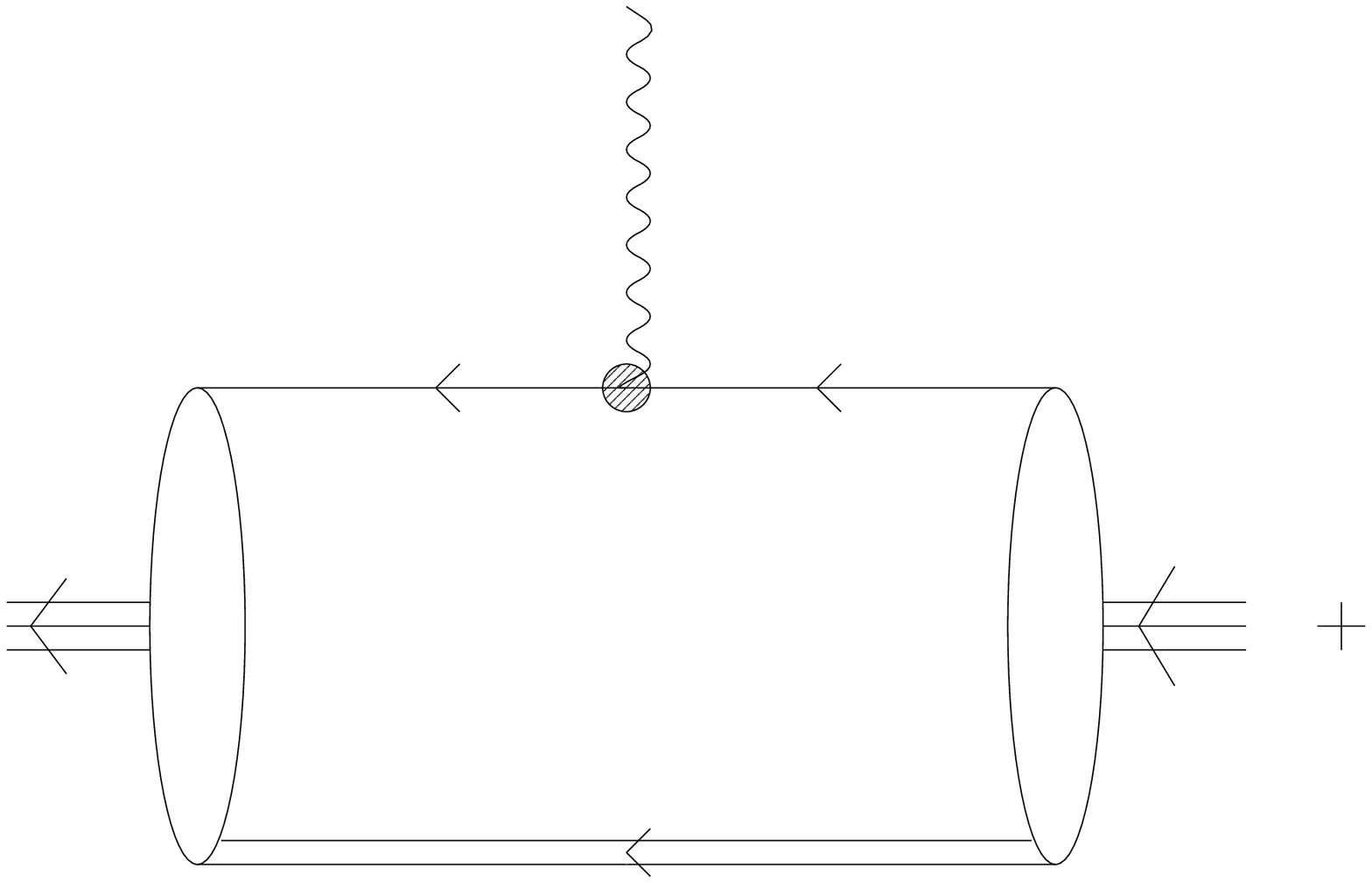}}{
\epsfxsize 7.3cm
\epsfbox{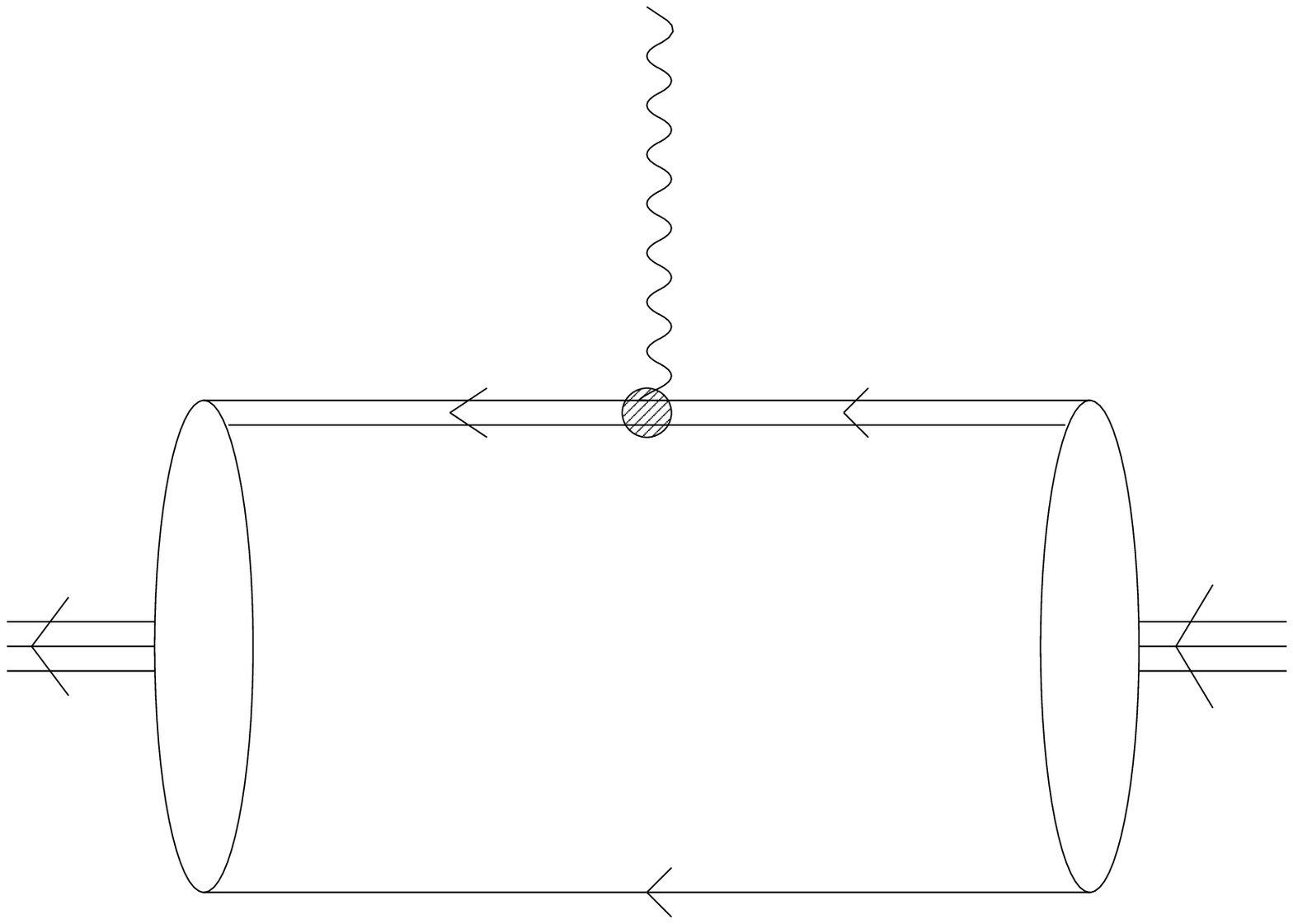}}}
%\end{center}
\caption{\label{impulsefig}The two diagrams which contribute to the nucleon 
matrix elements in impulse approximation.}
\end{figure}
\be
%\langle \hat O \rangle 
J_{(\mu)}(P_f,P_i) 
= \mintpf \mintpi
{\bar \chi} (P_f, p_f)D(\half P_f-p_f)S(\half P_f +p_f)\times
\nonumber\\*
\Gamma_{\hat O}(p_f,P_f;p_i,P_i) 
D(\half P_i-p_i)S(\half P_i+p_i) 
{\chi} (P_i,p_i).
\label{MANDELSTAM1}
\ee
This representation allows the direct evaluation of the nucleon current.
The 5-point function $\Gamma_{\hat O}$ appearing 
in eq. (\ref{MANDELSTAM}) and 
(\ref{MANDELSTAM1}) describes 
the coupling of the probing current with momentum $Q$ 
to the constituents of the bound state. 

In the following we work in a generalized impulse approximation, 
where the coupling to the diquark and the constituent quark 
is considered, while  we neglect, 
in this first calculation, 
the coupling to the exchanged 
quark.
\\
Accordingly the 5-point function is defined by 
\be
\Gamma_{\hat O}(p_f,P_f;p_i,P_i)= 
\delta(p_f-p_i-\half Q)D^{-1}(\half P_i-p_i)
\Gamma_{\hat O}^{q}(p_f,P_f;p_i,P_i) \nonumber\\*
+\delta(p_f-p_i+\half Q)S^{-1}(\half P_i+p_i)
\Gamma_{\hat O}^{d}(p_f,P_f;p_i,P_i).
\ee
Inserting $\Gamma_{\hat O}$ into eq. (\ref{MANDELSTAM1}), 
one observes that 
the matrix element is split into a quark and a 
diquark part
\be
%\langle \hat O \rangle = 
J_{(\mu)}(Q^2) = J_{(\mu)}^q(Q^2) + J_{(\mu)}^d(Q^2).
\label{nuccurr}
\ee
If the external nucleon legs are on their mass shells ($P_i^2=P_f^2=-M^2$),
the nucleon current  
depends only on the squared momentum $Q^2$ of the probing external current.
The  two terms appearing in (\ref{nuccurr}) are given by the loop integrals 
\be
J_{(\mu)}^q(Q^2)=\mintp \bar \chi(P_f,p_f)S(p_+)\Gamma_{(\mu)}^{q}(p_+,
p_-)S(p_-)D(p_d)\chi(P_i,p_i),
\label{quarkpart}
\ee
and 
\be
J_{(\mu)}^d(Q^2)=\mintp \bar \chi(P_f,p_f)D(p_+)\Gamma_{(\mu)}^{d}(p_+,
p_-)D(p_-)S(p_q)\chi(P_i,p_i).
\label{diquarkpart}
\ee
In appendix \ref{loop} we discuss the momentum rooting in the loop integrals 
and their numerical evaluation.
\\
Since the Bethe-Salpeter 
equation does not  provide the  overall normalization of the bound state 
vertex function, a physical 
normalization condition has to be imposed. The usual requirement,
a unit residue of the bound state propagator at the mass pole \cite{Itz85}, 
which in our formalism reads
\be
\mintp \bar \chi(P,p)
[\frac{\del}{\del P_\mu}S(\half P +p)D(\half P-p)]
\chi(P,p)
 = 2 \Lambda^+ P^\mu,
\label{BSENORM}
\ee
then leads to the correct normalization.
Since the quark exchange is independent of the c.m. momentum
$P$, it does not  contribute to the normalization of the vertex function.
Note that the adjoint vertex function  $\bar \chi (P,p)$ has to fulfill
\be
\bar \chi(P,p)= \Lambda^+ \bar \chi(P,p) 
\ee
and is therefore given by 
$\bar \chi(P,p)=\gamma_4 \chi^\dagger(P,p)\gamma_4$. 

\subsection{Electromagnetic Form Factors}
To determine the  electromagnetic (e.m.) form factors
one has to  calculate the e.m. nucleon current.
The probing current is an external photon
and therefore 
the  quark-photon and the diquark-photon vertex functions are  
needed to apply Mandelstam's formalism.
Electromagnetic gauge invariance is ma\-ni\-fest 
in  the Ward-Takahashi identity,
\be
Q_\mu \Gamma_\mu(p+Q,p) = S^{-1}(p+Q)-S^{-1}(p),
\label{WTI}
\ee
which provides the  connection between the 
longitudinal part of the (di-)quark-photon vertexfunction 
and the inverse (di-)quark propagator,
and for the limit $Q^2 \rightarrow 0$ in the 
Ward identity
\be
\Gamma_\mu(p,p)=\frac{\del}{\del p_\mu} S^{-1}(p).
\label{WI}
\ee
Using vertex functions in accordance with these identities
means to respect gauge invariance at the constituent level.
\\
A vertex function which solves the above identites for the quark-photon
coupling is the 
Ball-Chiu vertex \cite{Bal80,Kus83},
\be
(\Gamma_\mu^{q})_{e.m.}(p,k)&=&
\Gamma_\mu^{BC}(p,k)=i\frac{A(p^2)+A(k^2)}{2}\gamma_\mu
\nonumber\\*
&+& i \frac{(p+k)_\mu}{p^2-k^2} \left[(A(p^2)-A(k^2))\frac{\gamma p 
+ \gamma k}{2} 
-i(B(p^2)-B(k^2))\right],
\label{BCV}
\ee
which is determined by the quark self-energy functions 
$A(p^2)$ and $B(p^2)$ appearing in the quark propagator
\be
S(p)=-i\gamma\!\cdot\! p \sigma_v(p^2)+\sigma_s(p^2)
=[i\gamma\!\cdot\! p A(p^2) +B(p^2)]^{-1}
\label{quarkp}
\ee
and can be read off from eq. (\ref{S}).
The vertex function (\ref{BCV})
not only satisfies eqs. (\ref{WTI}) and (\ref{WI}), but is also
free of kinematical singularities, has the same properties under 
P,C,T transformations as the perturbative vertex $i\gamma_\mu$ and reduces
itself to this vertex in case of bare quarks (this is the case when 
a type I quark propagator is used).
In the following we will restrict ourselves to the pure 
longitudinal Ball-Chiu vertex and will 
not consider any transversal part, which cannot be determined
using  e.m. gauge invariance\footnote{In ref. \cite{Cur92} it was 
shown that the requirement of multiplicative renormalisibility 
constrains the construction
of possible transversal parts of the e.m. vertexfunction.}.
\\
The vertex function of a nonperturbative (extended) scalar particle,
also has to be chosen
to fulfill  eqs. (\ref{WTI}) and (\ref{WI}); this 
is discussed in ref. \cite{Oht90}. Again we will use 
the simplest longitudinal diquark-photon vertexfunction   
\be
(\Gamma_\mu^{d})_{e.m.}(p,k)=-(p+k)_\mu 
%\frac{D^{-1}(p)-D^{-1}(k)}{p^2-k^2},
\left[\frac{p^2 C(p^2)-k^2 C(k^2)}{p^2 -k^2}- m_{s}\frac{C(p^2)-C(k^2)}
{p^2-k^2}\right]
\ee
which is determined by the self-energy function $C(p^2)$ of the diquark 
propagator 
\be
D(p) = -\frac{F(p^2)}{p^2+m_s^2}=-\left((p^2+m_s^2)C(p^2)\right)^{-1}.
\ee
The explicit expression can be easily obtained 
when comparing  with eq. (\ref{Ds}).
In particular, any transversal part of the vertex function, 
which would describe an anomalous magnetic moment of the scalar diquark
will not be considered.
\\
To calculate the electromagnetic form factors of the nucleon 
we insert
\be
\Gamma^q_\mu(p_+,p_-)= Q_q (\Gamma_\mu^{q})_{e.m.}(p_+,p_-)
\ee
and 
\be
\Gamma^d_\mu(p_+,p_-)= Q_d (\Gamma_\mu^{d})_{e.m.}(p_+,p_-)
\ee
with the charge matrices in isospin space given by 
\be
Q_q = \half(\third {\bf 1} +\mbox{\bf $\tau$}_z)
,  \quad Q_d = \third.
\ee
into eqs. (\ref{quarkpart}) and (\ref{diquarkpart}), respectively, and
evaluate the loop integrals.
As it is well known, using Lorentz invariance, invariance under P,C,T
and the fact that the incoming and outgoing nucleons are onshell,
the longitudinal e.m. nucleon current can be decomposed
into
\be
J_\mu^{e.m.}(Q^2)
% &=& \langle N(P_f,S_f)|j_\mu|N(P_i,S_i) \rangle
%\nonumber\\*
&=&
\Lambda^+(P_f,S_f)[iM_B(F_e(Q^2)-F_m(Q^2))\frac{P_\mu}{P^2}
+F_m(Q^2)\gamma_\mu]
\Lambda^+(P_i,S_i).\nonumber\\*
\ee
Note that within our formalism the Dirac spinors usually 
appearing on the right hand 
side of this equation are replaced by 
the $\Lambda^+$ projectors.
The Lorentz invariant functions 
$F_e(Q^2)$ and $F_m(Q^2)$
denote the electric and magnetic form factor, which
can be extracted by taking appropriate traces.
Here and in the following subsections the calculations 
are most conveniently performed in the Breit-frame (see appendix 
\ref{loop}).

\subsection{Pion Nucleon Form Factor}
The pion nucleon form factor, which enters into pure hadronic models 
usually as an input parameter, has been much debated in the last few 
years. Especially the value at the pion mass shell is not directly
accessible in experiments. For a recent analysis obtained from 
different experiments ($\pi N$ scattering, $NN$ scattering, ...)
see ref. 
\cite{Eri93}.
In this respect 
there is certainly the need
to calculate this observable from a more fundamental quark level.   
\\
Within our approach the pion nucleon form factor  
is obtained by coupling an external 
pion current to the diquark-quark bound state using again Mandelstam's
formalism.
Because of parity conservation one notes that 
there is no contribution from the 
loop integral (\ref{diquarkpart}) where the pion would couple to 
the $0^+$ diquark. So the relevant nucleon current is completely
carried by the quark part (\ref{quarkpart}).
In order to specify the vertex function for the pion-quark coupling
we use spontaneously broken chiral symmetry 
and Goldstone's theorem \cite{Del79}: In the chiral limit
the vertex function, which is nothing else than the pion Bethe-Salpeter
amplitude, is proportional to the scalar self energy of the quark.
This is a consequence of the fact that in the chiral limit (vanishing
quark current mass) where, provided the pion mass vanishes (Goldstone's 
theorem), the
pion Bethe-Salpeter equation is equivalent to the quark Dyson-Schwinger
equation. 
Therefore the proper choice is
\be
\Gamma_{\pi}^a(P_{\pi}^2=Q^2=0,p^2)=i\gamma_5 \frac{B(p^2)}{f_\pi}\tau^a, 
\label{PIONBSE}
\ee
where $B(p^2)$ is defined in eq. (\ref{quarkp}) and 
the experimental pion decay constant $f_\pi=0.093$ GeV is used.
Although this choice of the pion-quark vertex is 
valid only at the pion mass shell $Q^2=0$, we assume that the offshell
amplitude does not vary too much with the pion momentum and 
therefore the most obvious 
generalization \cite{Rob94}
\be
\Gamma_{\pi}^a(Q^2,p^2)=i\gamma_5 \frac{\tau^a}{2} \frac{1}{f_\pi} 
( B((p-\half Q)^2)	
+B((p+\half Q)^2))    
\ee
of the onshell vertex function is allowed at least for a few hundert 
MeV around the onshell point.  
\\
When inserting the pion quark vertex function into eq. (\ref{quarkpart})
and evaluating the loop integral we determine the pion-nucleon form factor.
Furthermore since the pion-nucleon matrix element is parametrized as
\be
J_{\pi}(Q^2) &=& \langle N(P_f,S_f)|j_{\pi}(Q^2)|N(P_i,S_i) \rangle
\nonumber\\
&=&
\Lambda^+(P_f,S_f)[\gamma_5 g_{\pi NN}(Q^2)\tau^a]\Lambda^+(P_i,S_i).
\ee
the extraction of $g_{\pi NN}(Q^2)$ is straightforward. 

\subsection{Axial Form Factor}
Finally we calculate the axial form factor $g_A(Q^2)$ of the nucleon.
Due to parity, we again have to consider only the quark part of the
nucleon current, since an axial current does not couple 
to the scalar diquark. To specify the relevant vertex function, where the
axial current interacts with the quark, we use the 
chiral Ward identity which reads
\be
Q_\mu \Gamma_{5 \mu}(p+Q,p) = S^{-1}(p+Q)\gamma_5 -\gamma_5 S^{-1}(p).
\label{AWTI}
\ee
This symmetry constraint can  be satisfied with  the vertex \cite{Del79}
\be
\Gamma_{5 \mu}^{a}(p,k)&=&
\left(i \frac{A(p^2)+A(k^2)}{2}\gamma_\mu
\right.
\nonumber\\*
&+& \left.
i\frac{(p+k)_\mu}{p^2-k^2} \left[(A(p^2)-A(k^2))\frac{\gamma p + \gamma k}{2} 
-i(B(p^2)+B(k^2))\right]\right)\gamma_5 \frac{\tau^a}{2} 
,
\label{ABCV}
\ee  
that is totally determined by the quark self-energies.
Although the axialvector  vertex function 
looks, up to the isospin generators and 
the explicit $\gamma_5$, very similar to the Ball-Chiu vertex (\ref{BCV})
(used in the case of a vector current coupling), 
there is one important difference: In the 
last term the scalar quark self-energies $B(p^2)$ and $B(k^2)$ add up, 
which in the limit $Q=p-k \rightarrow 0$ leads to
\be
Q_\mu \Gamma_{5 \mu}^a (p,k) \longrightarrow  \tau^a B(p^2) \gamma_5. 
\label{GAMMALIMES}
\ee
When comparing (\ref{GAMMALIMES})
with eq. (\ref{PIONBSE}), the pion-quark vertex, one observes the 
exact Goldberger Treiman relation for the quarks in the chiral limit:
\be
\lim_{Q \rightarrow 0} i Q_\mu \Gamma_{5 \mu}^a(p,k)
=f_\pi \Gamma_{\pi}^a(0,p^2). 
\ee
Stated in other words, in the chiral limit the axialvector quark 
coupling for vanishing momentum transfer is completely 
dominated by the pseudoscalar coupling to a massless pion
(conserved axial current).
\\
To examine  whether the phenomenologically  much more important 
Goldberger-Treiman relation on the nucleon level
\be
g_{\pi NN} = g_A \frac{M}{f_\pi}
\label{GTR}
\ee
is fulfilled, we have to determine $g_A(Q^2)$
by taking the matrix element of the axial current between the diquark-quark
bound states. 
After using the axialvector vertex function (\ref{ABCV})
to evaluate (\ref{quarkpart})
and comparing the result with the familiar decomposition 
\be
J_\mu^a(Q^2) &=& \langle N(P_f,S_f)|j_\mu^a(Q^2)|N(P_i,S) \rangle
\nonumber\\*
&=&
\Lambda^+ (P_f,S_f)\frac{\tau^a}{2}[\gamma_\mu g_A(Q^2)
+Q_\mu g_P(Q^2)]\gamma_5 \Lambda^+ (P_i,S_i) \nonumber\\*
\ee
of the matrix element, the invariant form factors are
immediately accessible. 
Note that besides $g_A(Q^2)$, the axial form factor 
of the nucleon, there exist another form factor $g_P(Q^2)$, the induced
pseudoscalar form factor with a pole at the pion mass shell.

\subsection{Discussion of Numerical Results} 
\label{formres}
\begin{figure}[h]
\vspace{0.3cm}
\centerline{{
\epsfxsize 8.2cm
\epsfbox{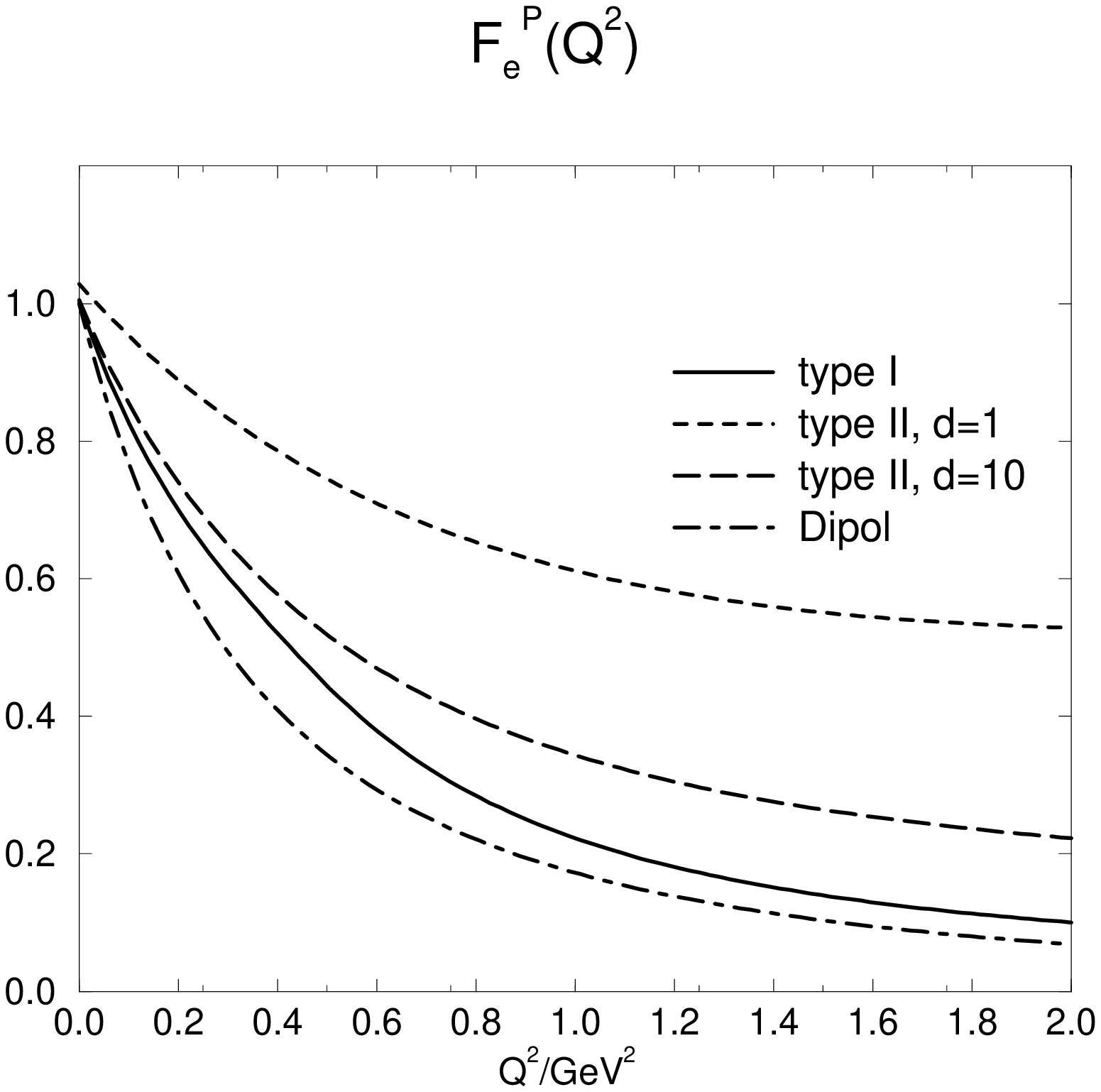}}{
\epsfxsize 8.2cm
\epsfbox{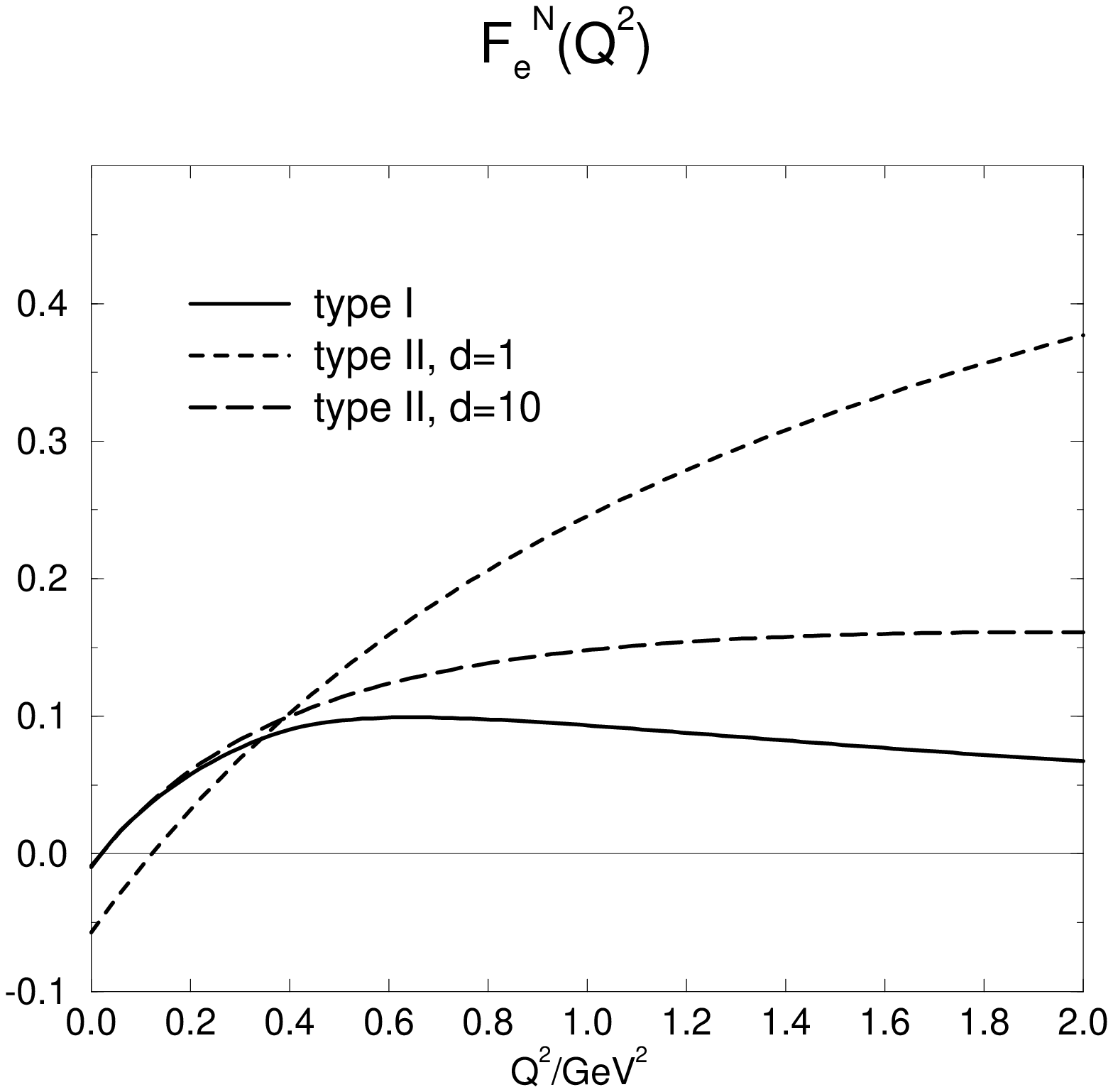}}}
%\end{figure}
%
\vspace{0.3cm}
%
%\begin{figure}[h]
\centerline{{
\epsfxsize 8.2cm
\epsfbox{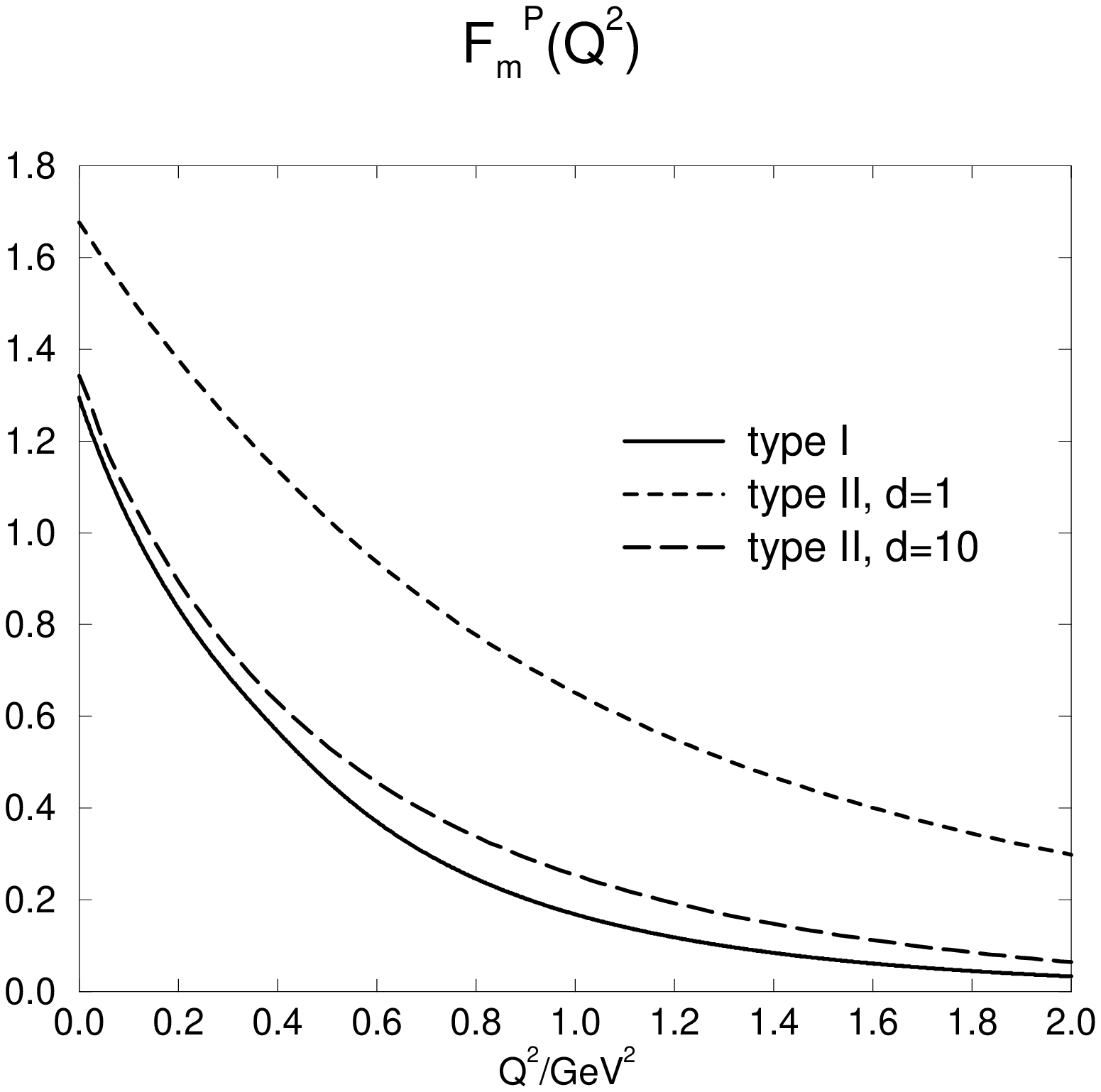}}{
\epsfxsize 8.2cm
\epsfbox{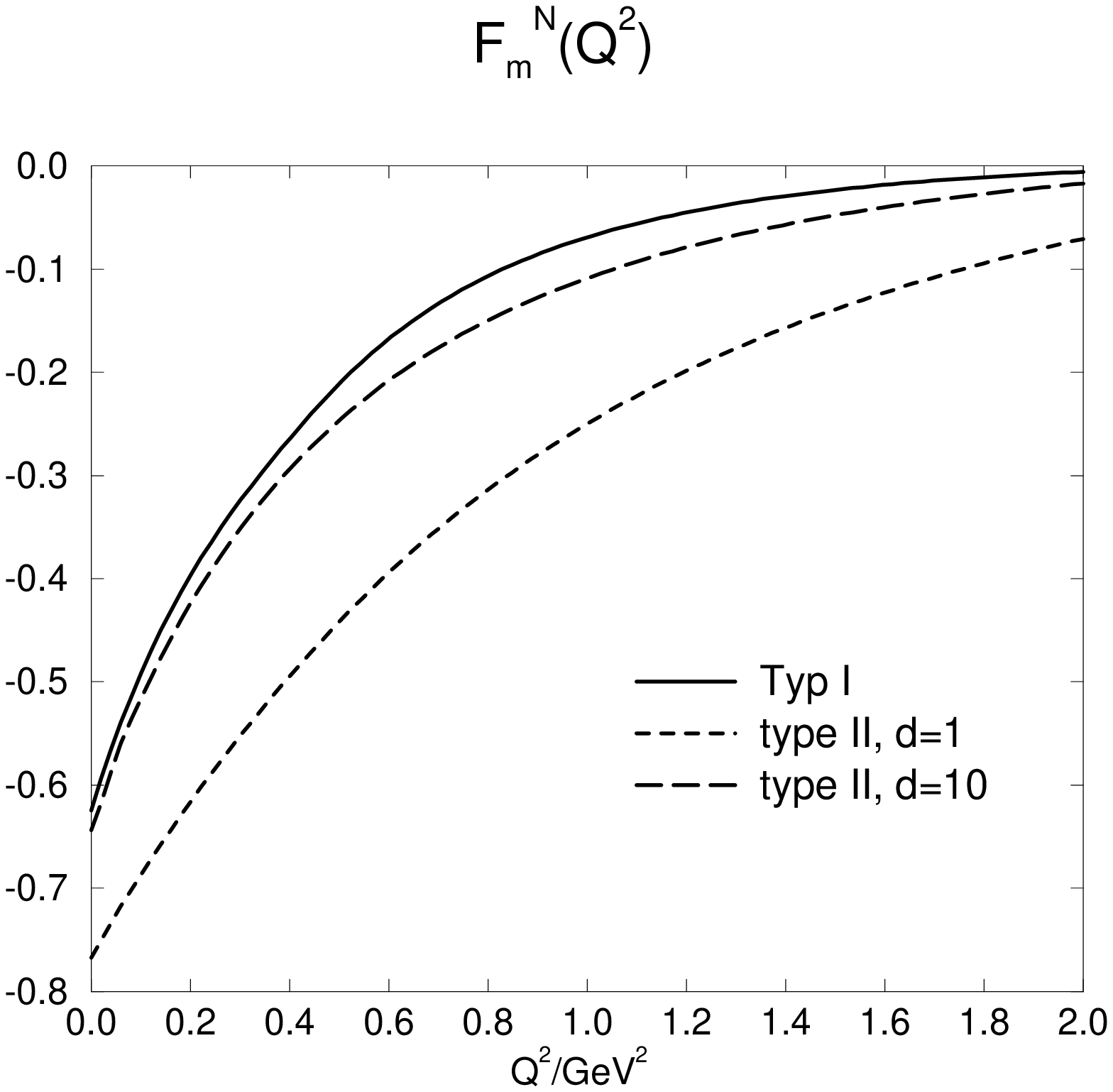}}}
\caption{Electromagnetic form factors of the nucleon.The nucleon 
mass in all cases is chosen to be
$M=1.9m_q$ and the amplitudes entering the
calculation of the matrix elements are obtained with $\Lambda=2m_q$.
\label{emform}}  
\end{figure}

In this subsection we finally show and discuss our results for the
nucleon form factors. We will concentrate again on three different 
cases: Using type I propagators, the solution of the corresponding 
Bethe-Salpeter equation and the appropriate vertex functions;
using type II propagators with $d=1$ and with $d=10$.
The so obtained electromagnetic form factors of the nucleon are shown in
fig. (\ref{emform}). 

First of all, we mention (as  can also be seen in the figures), 
that it is not possible in general to obtain exact charge conservation.
This is due to the impulse approximation, where the coupling of the photon 
to the exchanged quark (and also to the diquark-quark form factor)
is neglected. Whereas in a Salpeter approach 
charge conservation can be shown 
analytically \cite{Kei96b}\footnote{Going to the 
Salpeter limit we also get charge conservation
numerically.}
and it is well known that the impulse approximation is sufficient 
to guarantee charge conservation of mesons within 
the Bethe-Salpeter approach \cite{Rob96}, in our fully relativistic
nucleon calculation the situation is more complicated  
(see also ref. \cite{Asa95}). But 
it can also be seen in the figures that the violation of charge 
conservation $\Delta Q_{P,N}$ is at the most 
only a few percent (due to isospin 
symmetry $\Delta Q_N=-2\Delta Q_P$). In this respect the contribution 
of the neglected exchange graph is more important for the case of a
minimal screening of the mass poles (type II with $d=1$).  

Furthermore  the electric form factor of the 
proton in calculations of type I and type II
with $d=10$ are in reasonable correspondence with the empirical dipol fit
\be
F_e^P(Q^2)=\frac{1}{(1+Q^2/0.71 {\rm GeV}^2)^2}. 
\ee
The result obtained with $d=1$ falls short in this respect: 
The variation of $F_e^P(Q^2)$
with the photon momentum is too weak. This feature is also present 
for the other form factors. Obviously the modifications of the propagators
with a pure exponentional factor severly changes the spacelike
observable nucleon properties.

The electric neutron form factor shows a reasonable course. 
Note that a sensible description
of neutron properties was one of the motivations for the diquark concept 
\cite{Dzi81}, having the following intuitive (nonrelativistic)
picture in mind: A lighter quark circling around a heavier scalar diquark
inevitably leads to a negative mean square radius of the neutron.
It is now  satisfying to observe that our relativistic approach,
although working with equal masses of the constituents, confirmes  
this intuitive picture; it is not trivial at all, that the quark and 
diquark contributions give rise to a qualitatively correct 
behaviour of the electric neutron form factor.

\begin{table}[h]
\vspace{0.2cm}
\centering
\begin{tabular}{||c||c|c|c|c||}
\hline
    \multicolumn{5}{||c||}{\rm Nucleon form factors} \\
\hline
        & type I       & type II, $d=1$  & type II, $d=10$   &  Exp.   \\ 
\hline
  $Q_P$ &$1+4.1\cdot 10^{-3}$ &$1+2.85\cdot 10^{-2}$&$1+4.7\cdot 10^{-3}$&
 1          \\ 
  $Q_N$ &$-8.2 \cdot 10^{-3}$&$-5.70\cdot 10^{-2}$&$9.4\cdot 
 10^{-3}$& 0          \\ 
\hline
$\mu_P$ &  1.32    &  1.68           & 1.34              & 2.79   \\ 
$\mu_N$ &  -0.64   &  -0.77          & -0.65             & -1.91  \\ 
\hline
$g_A$   &  1.39    &   1.31          & 1.41              & 1.25         \\ 
$g_{\pi NN}$ &  10.59  &  14.17      &  10.89            & 13-14.5 \\ 
%\hline
%$\tilde g_A$   &          &                 &                   &          \\ 
%$\tilde f_{\pi NN}$ &          &                 &                   &   \\ 
\hline
\hline
\end{tabular}
\\
\label{formff}
\caption{In this table we show the basic results for the form factors. Note
that all magnetic moments are given in in units of $e/2M$. All displayed
observables are obtained in calculations, where $S_1(p)$ and $S_2(p)$
were included. The number of considered Gegenbauer polynomials was 
terminated after convergence was achieved.}
\end{table}
%\vspace{2cm}
%

%
\begin{figure}[h]
\vspace{0.8cm}
\centerline{{
\epsfxsize 8.2cm
\epsfbox{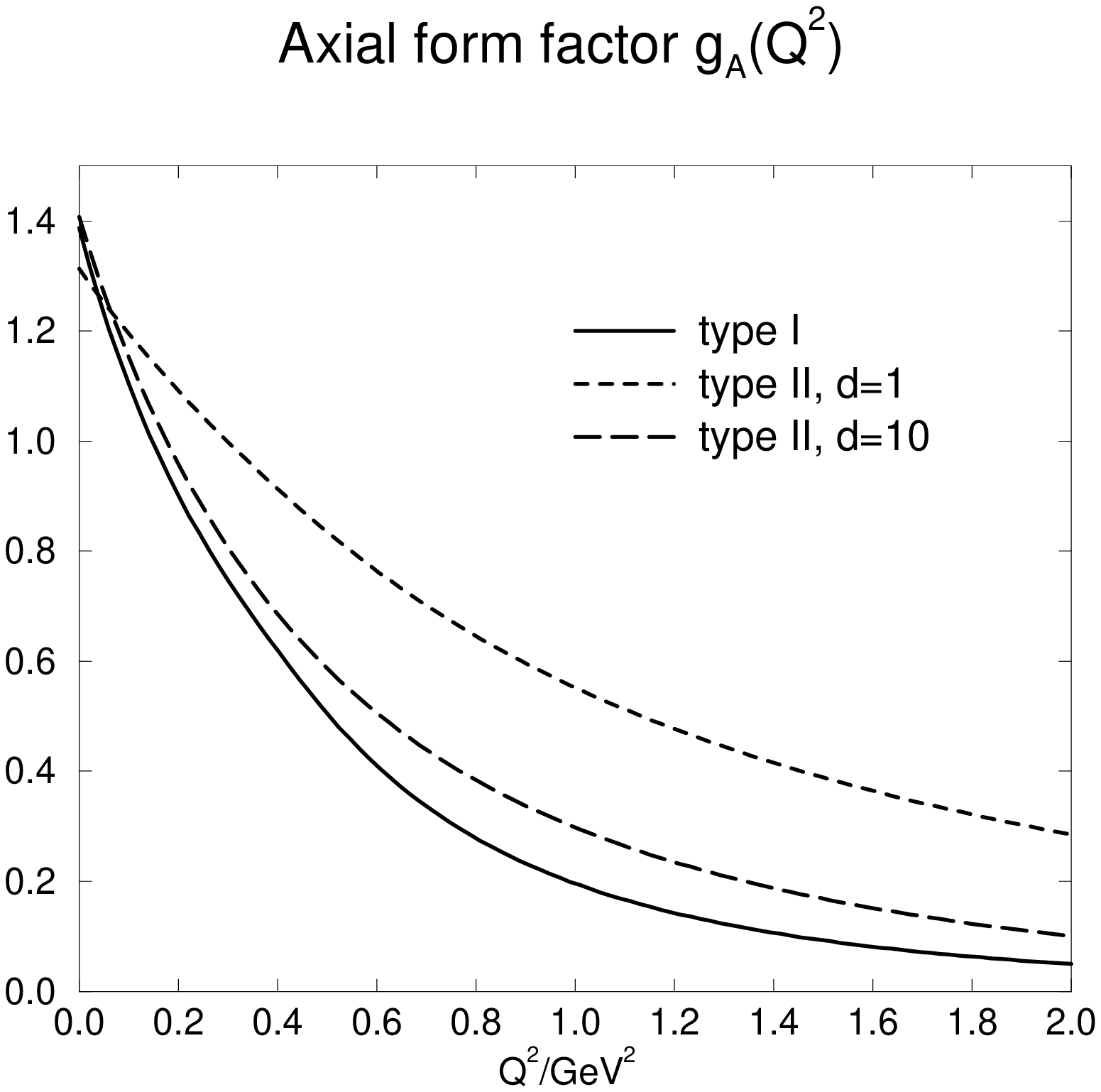}}{
\epsfxsize 8.2cm
\epsfbox{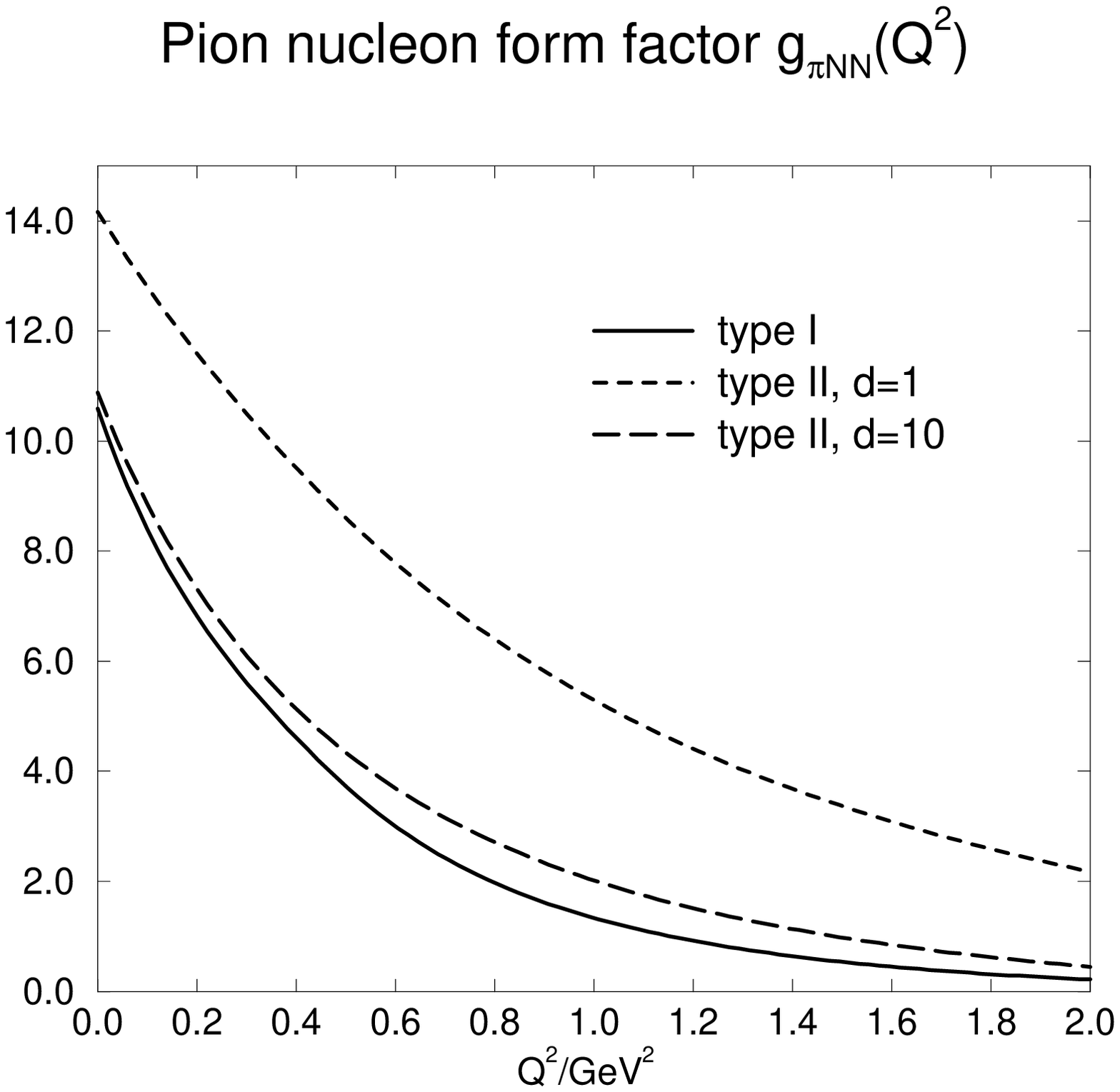}}}
\caption{The axial form factor $g_A(Q^2)$ of the nucleon and the 
pion nucleon form factor $g_{\pi NN}(Q^2)$ are shown for  type I
and  type II with $d=1, 10$ calculations respectively.\label{axfig1}}  
%\end{figure}
%\begin{figure}[b]
\vspace{1.cm}
\centerline{{
\epsfxsize 8.2cm
\epsfbox{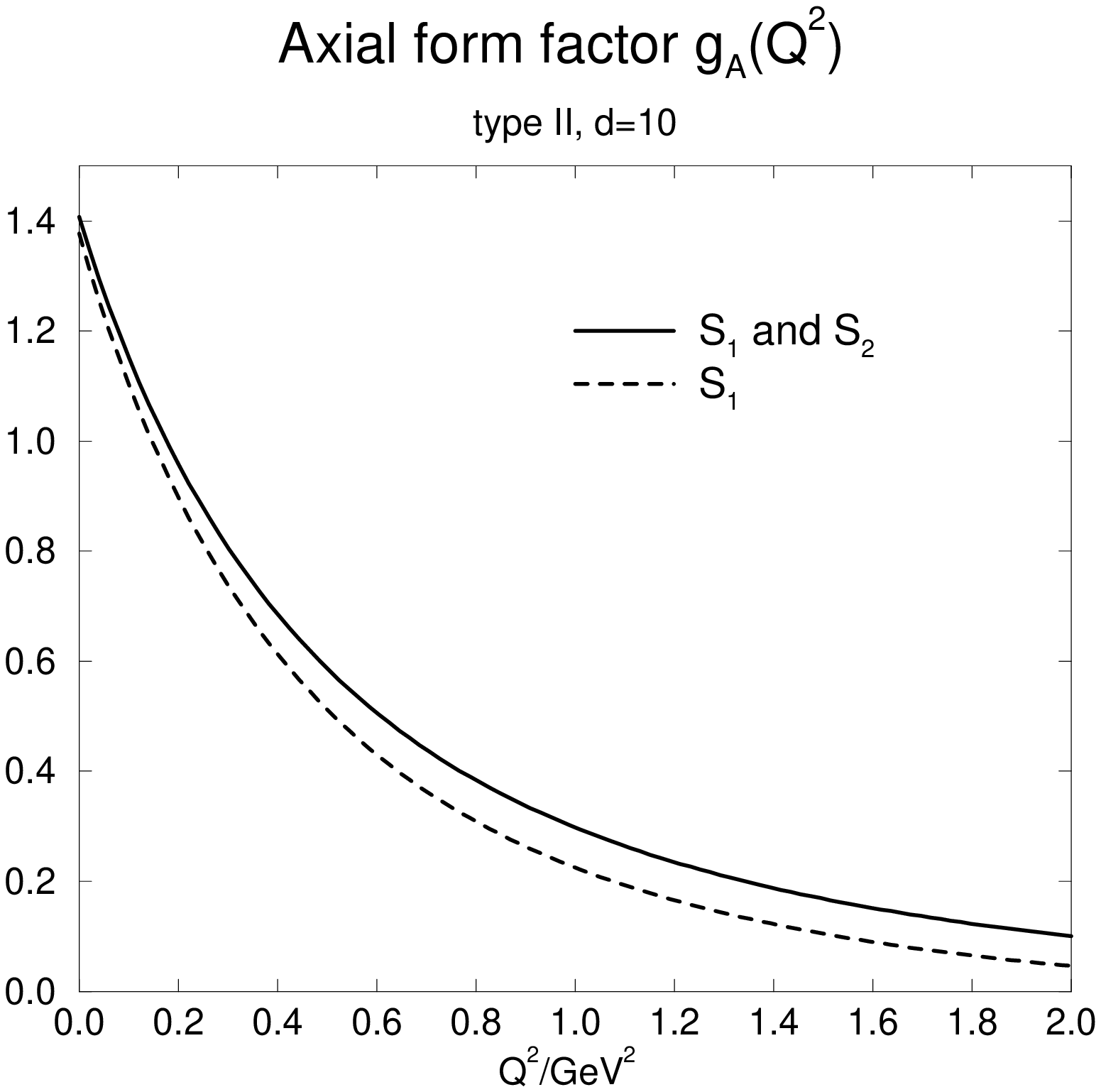}}{
\epsfxsize 8.2cm
\epsfbox{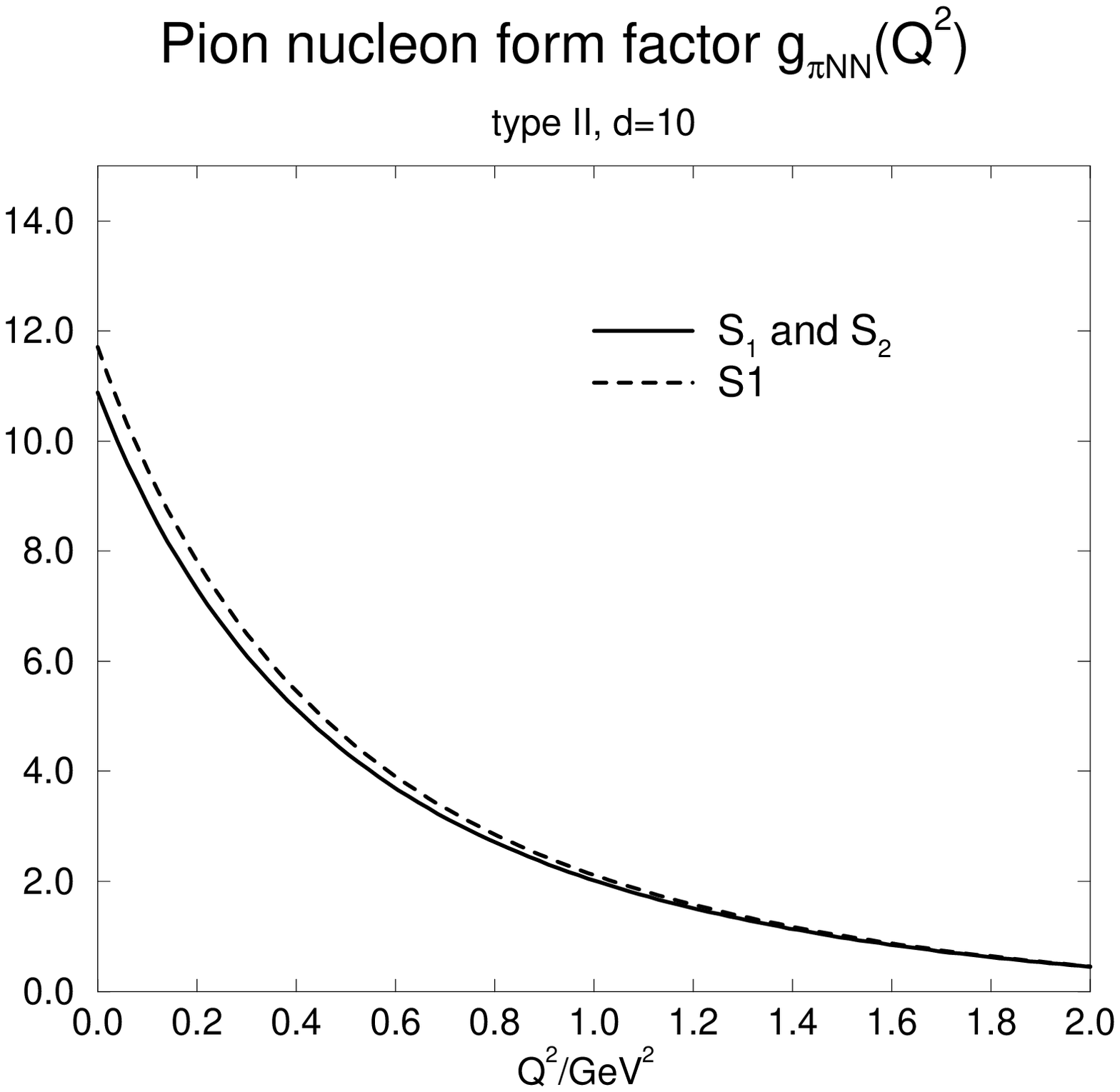}}}
\caption{The axial form factor $g_A(Q^2)$ and the 
pion nucleon form factor $g_{\pi NN}(Q^2)$ are shown 
for a type II with $d=10$ calculation, including only $S_1(p)$ or 
including $S_1(p)$and $S_2(p)$.\label{axfig2}}  
\end{figure}

For both proton and neutron, the magnetic moments are too small. 
This is, however, not surprising because large contributions 
coming  from $1^+$ diquarks
are  expected for these observables (\cite{Kei96b}, \cite{Wei93}).
Nevertheless, the calculated values have magnitudes expected from 
a calculation involving only $0^+$ diquarks 
\cite{Hel95,Kei96a}.

Before discussing the axial properties of the nucleon, we mention 
the following behaviour of the form factors: When $\Lambda$ is decreased,
leading to narrower amplitudes in momentum space 
(compare fig. (\ref{width})), the electric 
form factors for the type I calculation become steeper and approach
the emperical dipol fit. While such a behaviour is expected near threshold
\cite{Jaf92} we do not observe this feature for type II calculations.
In these calculations the form factors (at leat at small $Q^2$) are 
very insensitive on $\Lambda$ and the actual value of the bound state
mass.

Our results for the axial  and the pion-nucleon 
form factor are shown in fig. (\ref{axfig1}). Again, we observe for
a type II with $d=1$ calculation a relatively small variation 
of the form factors with $Q^2$. Furthermore for the calculation including 
type II propagators with $d=10$ the similarities to the type I case
can be seen.
A striking feature, however, is the good agreement of $g_A=g_A(Q^2=0)$ and
$g_{\pi NN}=g_{\pi NN}(Q^2=0)$ with the experimental values. $g_A$ 
tends to be slightly too large 
whereas the pion nucleon coupling constant
obtained with type I and with type II (d=10) is a little bit too small. 
Nevertheless, in all cases a qualitatively right behaviour is
obtained. This is astonishing, because also for these observables
(in analogy to the magnetic moments) sizeable contributions 
from $1^+$ diquarks are expected. Note that in 
ref. \cite{Asa95} also a relatively 
large value of $g_A$ has been obtained, including only 
$0^+$ diquarks. 
 
By comparing the calculated values of $g_{\pi NN}$ with the ones obtained 
form the Goldberger Treiman relation (\ref{GTR}) we 
observe a violation of the Goldberger Treiman relation
up to $30 \%$. This failure also signals the necessity 
for the inclusion of $1^+$ diquarks in the calculation. 
It is furthermore to be seen how much the exchanged quark contributes
to these observables.
   
In figure (\ref{axfig2}) we show the influence of the lower
component $S_2(p)$ of the nucleon Dirac field by comparing the results
obtained with all amplitudes with the result obtained with $S_1(p)$
alone. It is interesting to note that the values of $g_{\pi NN}(Q^2)$ 
at small values of the momentum transfer are influenced by 
$S_2(p)$; while the values of $g_A(Q^2)$ at larger values 
are slightly more affected. 
Nevertheless we conclude that having a large lower component is
probably only due to the representation chosen for the Dirac matrices.
They do not have a significant effect on observables studied so far. 

\section{Conclusions and Outlook}
\label{Conclusions}
In this paper we have developed a covariant diquark-quark model suitable
for the calculation of nucleon observables. As discussed, confinement
is put in by a modification of the free (tree-level) quark and 
diquark propagators. The structure of the Bethe-Salpeter equation,
describing nucleons as diquark-quark bound states interacting through 
quark exchange, has been  elaborated
using an appropriate decomposition of the Bethe-Salpeter 
vertex function in the Dirac algebra. Furthermore, the coefficient 
functions of the corresponding Lorentz tensors have been expanded
in hyperspherical harmonics, exploiting an approximate $O(4)$
symmetry (which would be exact if the exchanged particle were massless).
After discussing the various  numerical methods we presented 
our results for the calculation including $0^+$ diquarks.
The advantages of the confining propagators are clearly seen in the
absence of unphysical thresholds.

The numerically obtained nucleon vertex 
function has been  fitted to a simple analytic form and then used
to calculate nucleon
matrix elements and  form factors. In particular,
we considered electromagnetic, the axial and the pionic form factors of the
nucleon. While, due to the various simplifications 
and approximations, a quantitative agreement of the results with experimental
values could not be expected, we nevertheless got reasonable results.
Furthermore, the large lower component of the nucleon
Dirac field was shown to be merely an artefact of the chosen 
spinor representation. Although, the various observables 
displayed that the developed picture of a nucleon is still
oversimplified, they  nevertheless give us confidence that 
we are on the right track in obtaining a reasonable and trustable 
baryon model. The next step of improvement will be the inclusion of $1^+$
diquarks in the Bethe-Salpeter equation and also in the nucleon 
matrix elements. Furthermore we plan to determine 
the internal structure of 
the diquarks (described crudely by the parameter $\Lambda$)
by a microscopic diquark model. 

In order to  fully exploit the advantages and possibilities of our covariant 
and confining approach, the model will soon be applied to other 
processes. In particular, we will investigate 
the different observables
associated with the reactions $p+\gamma \rightarrow K + \Lambda$ 
and $p+p \rightarrow K + \Lambda$ which are measured  
at ELSA \cite{Boc94} and COSY \cite{Eyr97}, respectively.\\
While in most calculations of nucleon structure functions 
the distribution functions of the constituents are 
merely parametrized (see \eg \cite{Mul97}), our approach 
offers the possibility to determine, in a first step, 
these distributions, after the constituents are specified.
In ref. \cite{Kus97} this was investigated, assuming the 
scalar diquark to be a spectator and the photon interacts only  
with the quark. 
 
We conclude, that the reported studies are a good starting point 
for further investigations, 
which certainly will improve our qualitative and quantitative
understanding of the baryon 
structure.

%\acknowledgments  
{\bf Acknowledgments }
We thank Rolf B\"aurle and Udo Z\"uckert for their 
contributions in the early stages of this work,
and G.\ Piller and K.\ Kusaka for the very helpful discussions on the 
numerical solution of the Bethe-Salpeter equation.
Helpful remarks by F.\ Lenz are gratefully acknowleged.
 
\appendix

\section{Dirac Decomposition}

We choose the following ansatz for the amplitude $\chi^\mu$
\be
\chi^\mu (P,p) &=&  A_1(P,p)\gamma_5\hat P^\mu \Xi \Lambda^+ 
+  A_2(P,p)  \gamma_5 \hat P^\mu \Lambda^+
\nonumber \\
&+&  \tilde B_1(P,p)\gamma_5\hat p^\mu_T \Xi \Lambda^+ 
 +   \tilde B_2(P,p)\gamma_5\hat p^\mu_T \Lambda^+
 \nonumber \\
 &+& \tilde C_1(P,p)  \gamma _5 i(\hat P^\mu \vect{1} -\gamma^\mu) \Lambda^+ 
  + \tilde C_2(P,p) \gamma _5 i(\hat P^\mu \vect{1} +\gamma^\mu) 
  \Xi \Lambda^+
  \label{Atilde}
\ee 
In the rest frame of the bound state, $P=(0,0,0,iM)$, we obtain by using
\be
\Lambda^+ = \pmatrix{1&0\cr 0&0\cr}, \quad 
\Xi\Lambda^+ =\pmatrix{0&0\cr\vect{{\hat p}}\vect{\sigma}&0\cr}.
\ee
the following expressions:
\be
\chi^4= \pmatrix{\vect{{\hat p}}\vect{\sigma} A_1(P,p) & 0 \cr
		\vect{1} A_2(P,p) &0 \cr}
\ee
and
\be
\vect{\chi} = \pmatrix{(\vect{{\hat p}}\vect{\sigma}) \vect{{\hat p}} 
\tilde B_1(P,p)
&0\cr       \vect{{\hat p}} \tilde B_2(P,p) &0\cr}
	  + \pmatrix{\vect{\sigma} \tilde C_1(P,p) & 0 \cr 
	 \vect{\sigma} (\vect{{\hat p}}\vect{\sigma}) \tilde C_2(P,p) 
& 0 \cr}
\ee

Using the redefinitions 
$$
\tilde B_i = B_i - C_i, \quad \tilde C_i = C_i
$$
turns eq.\ (\ref{Atilde}) into eq.\ (\ref{Adef}). In the rest frame this leads
to
\be
\vect{\chi} = \pmatrix{ \vect{{\hat p}} (\vect{{\hat p}}\vect{\sigma}) B_1(P,p)
		     + i (\vect{\sigma} \times \vect{{\hat p}} ) 
			 (\vect{{\hat p}}\vect{\sigma}) C_1(P,p) & 0 \cr
                     \vect{{\hat p}} B_2(P,p) + i (\vect{\sigma} 
              \times \vect{{\hat p}}) C_2(P,p) & 0 \cr}
\ee

Thus in the rest frame the meaning of the amplitudes is therefore as follows:
the amplitudes $A_i$ are the time components, the components $B_i$ are parallel
to the relative three--momentum of the constituents, and the 
components in the 
third line in
eq.\ (\ref{Adef}) are orthogonal to the momentum and the spin.

%\section{Hyperspherical Coordinates}
\section{Analytic fits of the amplitudes}
\label{fit}
\begin{figure}[h]
\centerline{{
\epsfxsize 8.5cm
\epsfbox{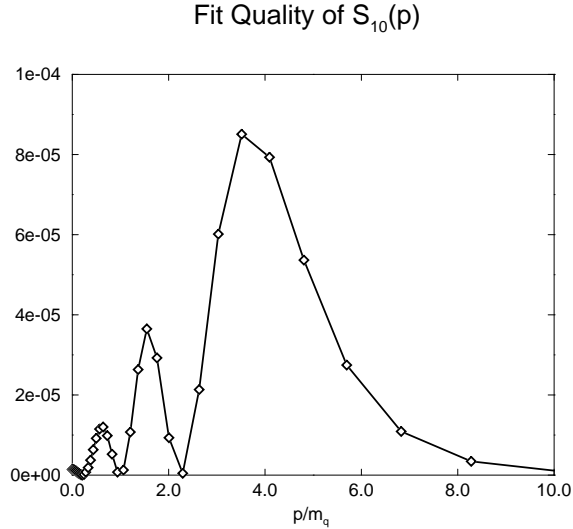}}}
\caption{
$(S_{10}(p)_{numeric} - S_{10}(p)_{fit})^2$ is plotted for the type II
calculation with $M=1.9 m_q$ \label{fitplot}.
} 
\end{figure}
\begin{table}[h]
\centering
\begin{tabular}{||c||c|c|c|c|c|c||}
\hline
    \multicolumn{7}{||c||}{\rm Fit coefficients for type I} \\
\hline
      &  $S_{10}$& $S_{11}$ & $S_{12}$ & $S_{20}$ & $S_{21}$ & $S_{22}$ \\ 
\hline
  $a^1$ &   0.9992 & -0.6954  &  0.03028 & -0.4191 &  0.4113  &  -0.0127 \\ 
  $b^1$ &   0.3107 &  0.3028  &  0.4617  & 0.2682  &  0.1937  &  0.4090  \\ 
  $a^2$ &  -0.0002 &  0.2929  &  0.0134  & -0.1841 & -0.3939  & -0.0058  \\ 
  $b^2$ &   7.8284 &  0.3028  &  0.1834  & 0.0792  &  0.1966  &  0.2085  \\ 
\hline
$\chi^2 \sigma$ & $6.786\cdot 10^{-5}$& $4.025\cdot 
10^{-5}$& $9.474\cdot
 10^{-9}$  
&$1.132\cdot 10^{-5}$ & $2.347\cdot 10^{-8}$& $1.207\cdot 10^{-10}$\\
\hline
\end{tabular}
\\
\vspace{0.3cm}
\caption{
The fit coefficients for a type I calculation
with $M=m_q$ and $\Lambda=2 m_q$ are displayed.
\label{fitI}}
\end{table}
\begin{table}[h]
\vspace{0.5cm}
\centering
\begin{tabular}{||c||c|c|c|c|c|c||}
\hline
    \multicolumn{7}{||c||}{\rm Fit coefficients for type II, $d=1$} \\
\hline
      &  $S_{10}$& $S_{11}$ & $S_{12}$ & $S_{20}$ & $S_{21}$ & $S_{22}$ \\ 
\hline
  $a^1$ &   1.6562 & -0.9674  &  0.4988  &-0.4168  &  0.3898  &  0.5042 \\ 
  $b^1$ &   0.2437 &  0.2556  &  0.1985  & 0.0924  &  0.1564  &  0.1562  \\ 
  $a^2$ &  -0.6574 &  0.6342  & -0.4686  & 0.0463  & -0.3528  & -0.5069  \\ 
  $b^2$ &   0.3838 &  0.3293  &  0.2046  & 0.0924  &  0.1613  &  0.1558  \\ 
\hline
$\chi^2 \sigma$ & $5.305\cdot 10^{-4}$& $1.340\cdot 10^{-4}$& $5.331\cdot
 10^{-7}$  
&$1.035\cdot 10^{-4}$ & $5.263 \cdot10^{-8}$& $3.594\cdot 10^{-9}$\\
\hline
\end{tabular}
\vspace{0.3cm}
\caption
{
The fit coefficients for a type II calculation
with $d=1$, $M=m_q$ and $\Lambda=2 m_q$ are displayed.
\label{fitII}
}
\end{table}
\vspace{1cm}
After solving the Bethe-Salpeter equation the amplitudes $S_{10}(p)$ ...
$S_{22}(p)$ are given only at the grid points of the momentum mesh. 
For further use
in the calculation of matrix elements 
this a very unconvenient feature. 
Therefore we fit  rational funtions to them.
To do this the value of every  amplitude 
at each grid point is therefore treated as a ``data'' point with 
a certain standart deviation $\sigma$.  
We find that the following set of rational functions are suitable
to take the behaviour at small $p$ as well as the behaviour at large
$p$ properly into account :
\be
S_{10}(p)&=&\sum_{k=1}^{2}\frac{a_{10}^k}{(1+b_{10}^k p^2)^2}\\  
S_{11}(p)&=&\sum_{k=1}^{2}\frac{a_{11}^k p}{(1+b_{11}^k p^2)^2}  \\
S_{12}(p)&=&\sum_{k=1}^{2}\frac{a_{12}^k p^2}{(1+b_{12}^k p^2)^3}  \\
S_{20}(p)&=&\sum_{k=1}^{2}\frac{a_{20}^k }{(1+b_{20}^k p^2)^3}   \\
S_{21}(p)&=&\sum_{k=1}^{2}\frac{a_{21}^k p^2}{(1+b_{21}^k p^2)^3}  \\
S_{22}(p)&=&\sum_{k=1}^{2}\frac{a_{22}^kp^3}{(1+b_{22}^k p^2)^4}  \\
\ee
Note, that the momentum $p$ is taken in units of the 
quark mass. 
The fit coefficients $a_{ij}^k, b_{ij}^k;\, i,j=1,2; \,k=1,2$ 
are then determined independently for each amplitude by performing 
a $\chi^2$ fit. To judge the quality of this procedure 
we show in fig. (\ref{fitplot}) the squared deviation
\be
\tilde \Delta^2 = (S_{10}(p)_{numeric}-S_{10}(p)_{fit})^2.
\ee
Suming up the squared deviations then leads to 
\be
\sigma^2 \chi^2 =\sum_{k=1}^{kmax} \Delta^2=\sum_{k=1}^{kmax} 
(S_{10}(k)_{numeric}-S_{10}(k)_{fit})^2,
\ee
i.e. the merit function $\chi^2$ times the squared accuracy 
of the ``data points''. 
In the tables (\ref{fitI}) and (\ref{fitII})
we show  the fit coefficients which reproduce the plots for the type I
and the type II with $d=1$ amplitudes given in
figure (\ref{allamp}) within very small deviations. 

\section{Evaluation of the loop integrals}
\label{loop}

In Section 4.1 we derived the two loop integrals building up 
the nucleon current. In eq. (\ref{quarkpart})
\be
J_\mu^q(Q^2)=\mintp \bar \chi(P_f,p_f)S(p_+)\Gamma_\mu^{q}(p_+,
p_-)S(p_-)D(p_d)\chi(P_i,p_i),
\ee
the momenta are definded as 
\be
p_{\pm} &=& p+\half P_i +\half Q \pm \half Q \nonumber\\*
p_d     &=& -p+\half P_i
\ee
whereas in eq. (\ref{diquarkpart})
\be
J_\mu^d(Q^2)=\mintp \bar \chi(P_f,p_f)D(p_+)\Gamma_\mu^{d}(p_+,
p_-)D(p_-)S(p_q)\chi(P_i,p_i).
\ee
the momenta are chosen as 
\be
p_{\pm} &=& -p+\half P_i +\half Q \pm \half Q \nonumber\\*
p_q     &=&  p+\half P_i.
\ee
As stated in the text, to evaluate these integrals it is useful
to work in the Breit-frame
\be
Q   &=& (0,0,Q_3,0) \nonumber\\*
P_i &=& (0,0,-\half Q_3,i\sqrt{M^2+\quarter Q^2}) \nonumber\\*
P_f &=& (0,0,\half Q_3,i\sqrt{M^2+\quarter Q^2}),
\label{BREITFRAME}
\ee
and further to express the loop momentum $p$ in 4-dimensional 
spherical coordinates
\be
p= p\left(\sin \phi \sin \theta \cos \psi,\sin \phi \sin \theta
\sin \psi,\sin \psi \cos \theta,\ \cos \psi\right).
\ee
While the $\phi$ integration is trivial, the integration over
$p$, $\Theta$ and $\psi$ has to be performed numerically.

\end{document}